\documentclass[a4paper]{article}
\usepackage[margin=2cm]{geometry}
\usepackage{graphicx}
\usepackage{amsmath}

\begin{document}

\title{From Artificial Neural Networks to Deep Learning\\
	for Music Generation
	-- History, Concepts and Trends\footnote{To appear in the Special Issue on Art, Sound and Design
		in the Neural Computing and Applications Journal.}}



\author{Jean-Pierre Briot${^\dagger}{^\ddagger}$}

\date{$^\dagger$ Sorbonne Universit\'e, CNRS, LIP6, F-75005 Paris, France\\
		$^\ddagger$ UNIRIO, Rio de Janeiro, RJ 22290-250, Brazil\\[.2cm]
		{\tt\small Jean-Pierre.Briot@lip6.fr}}

\maketitle

{\bf Abstract:}
The current
wave
of deep learning (the hyper-vitamined return of artificial neural networks)
applies not only to traditional statistical machine learning tasks: prediction and classification
(e.g., for weather prediction and pattern recognition),
but has already conquered other areas, such as translation.
A growing area of application is the generation of creative content,
notably the case of music, the topic of this paper.
The motivation is in using the capacity of modern deep learning techniques
to automatically learn musical styles from arbitrary musical corpora
and then to generate musical samples from the estimated distribution,
with some degree of control over the generation.
This paper provides
a tutorial on
music generation based on deep learning techniques.
After a short introduction to the topic illustrated by a recent exemple,
the paper analyzes some early works from the late 1980s using artificial neural networks for music generation
and how their pioneering contributions
have prefigured
current techniques.
Then, we introduce some conceptual framework to analyze the various concepts and dimensions involved.
Various examples of recent systems are introduced and analyzed to illustrate the variety of concerns and of techniques.

\section{Introduction}
\label{section:introduction}

Since the mid 2010s\footnote{In 2012, an image recognition competition
	(the ImageNet large scale visual recognition challenge)
	was won by a deep neural network algorithm named AlexNet \cite{Krizhevsky:2012:ICD:2999134.2999257},
	with a stunning margin over the other algorithms which were using handcrafted features.
	This striking victory was the event which ended the prevalent opinion
	that neural networks with many hidden layers could not be efficiently trained
	and which started the deep learning wave.},
deep learning has been producing striking successes and is now used routinely for classification and prediction
tasks, such as image recognition, voice recognition or translation.
It continues conquering new domains,
for instance
source separation\footnote{Audio source separation,
	often coined as the cocktail party effect, has been known for a long time to be a
	very difficult problem
	\cite{cherry:cocktail:problem:1953}.}
\cite{deep:learning:solves:cocktail:party:2015}
and text-to-speech synthesis \cite{oord:wavenet:arxiv:2016}.

A growing area of application of deep learning techniques is the generation of content,
notably music,
the focus of this paper.
The motivation is in using
widely available various musical corpora to automatically learn musical styles
and to generate new musical content based on
them.
Since a few years, there is a large number of scientific papers about deep learning architectures and experiments
to generate music, as witnessed in \cite{briot:dlt4mg:springer:2019}.
The objective of this paper is to explain some fundamentals as well as various achievements of this stream of research.

%
%

\subsection{Related Work and Organization}
\label{section:related:work}
\label{section:organization}

This paper takes some inspiration from
the
comprehensive survey and analysis
proposed by
the recent
book \cite{briot:dlt4mg:springer:2019},
but with a different organization and material
and it also includes an original historical retrospective analysis.
Another related article \cite{mgbdlcd:ncaa:2018} is an analysis focusing on
challenges.
In \cite{herremans:taxonomy:music:generation:acm:cs:2017},
Herremans {\em et al.} propose a function-oriented taxonomy for various kinds of music generation systems.
Some more general surveys about of AI-based methods for algorithmic music composition are
by Papadopoulos and Wiggins \cite{papadopoulos:ai:algorithmic:composition:1999}
and by Fern\'andez and Vico \cite{fernandez:ai:methods:algorithmic:composition:survey:jair:2013},
as well as books by Cope \cite{cope:algorithmic:composer:book:2000} and by Nierhaus \cite{nierhaus:algorithmic:composition:book:2009}.
In \cite{graves:generating:sequences:rnn:arxiv:2014},
Graves analyzes the application of recurrent neural networks architectures to generate sequences (text and music).
In \cite{fiebrink:ml:creative:tool:arxiv:2016},
Fiebrink and Caramiaux address the issue of using machine learning to generate creative music.
In \cite{pons:nn:music:history:web:2018},
Pons presents a short historical analysis of the use of neural networks for various types of music applications
(that we expand in depth).

This paper is organized as follows.
Section~\ref{section:introduction} (this section) introduces the general context of deep learning-based music generation
and includes a comparison to some related work.
Section~\ref{section:generation} introduces the principles and the various ways of generating music from models.
Section~\ref{section:first:example} presents
some introductory example.
Section~\ref{section:history}
analyzes in depth some pioneering works in neural networks-based music generation from the late 1980s and their
later
impact.
Section~\ref{section:framework} presents some conceptual framework
to
classify
various types of current deep learning-based music generation systems\footnote{Following the model
	introduced in \cite{briot:dlt4mg:springer:2019}.}.
We analyze possible types of representation in Section~\ref{section:representation}.
Then, we analyze successively:
basic types of architectures and strategies in Section~\ref{section:basic:architecture:strategy};
various ways to construct compound architectures in Section~\ref{section:architecture:compound};
and some more refined architectures and strategies in Section~\ref{section:refined:architecture:strategy}.
Finally, Section~\ref{section:open:issues} introduces some open issues and trends,
before concluding this paper.
A glossary annex completes the paper.

\section{Music Generation}
\label{section:generation}

In this paper, we will focus on {\em computer-based music composition}
(and not on computer-based sound generation).
This is often also named {\em algorithmic music composition}
\cite{nierhaus:algorithmic:composition:book:2009,cope:algorithmic:composer:book:2000},
in other words, using a formal process,
including steps (algorithm) and components, to compose music.

\subsection{Brief History}
\label{section:generation:history}

One of the first documented case of algorithmic music composition,
long before computers,
is the Musikalisches Wurfelspiel (Dice Music), attributed to
Mozart.
A musical piece is generated
by concatenating randomly selected (by throwing dices) predefined music segments
composed in a given style (Austrian waltz in a given key).

The first music generated by computer appeared in the late 1950s,
shortly after the invention of the first computer.
The Illiac Suite is the first score composed by a computer \cite{lejaren:illiac:book:1959}
and was an early example of algorithmic music composition,
making use of stochastic models (Markov chains) for generation,
as well as rules to filter generated material according to desired properties.
Note that, as opposed to the previous case which consists in rearranging predefined material,
{\em abstract models} (transitions and constraints) are used to guide the generation.

One important limitation is that the specification of such abstract models,
being rules, grammar, or automata, is difficult (reserved to experts) and error prone.
With the advent of machine learning techniques, it became natural to apply them to {\em learn} models from a corpus of existing music.
In addition, the method becomes, in principle, independent of a specific musical style\footnote{Actually,
	the style is defined {\em extensively} by (and learnt from) the various examples of music selected as the training examples.}
(e.g., classical, jazz, blues, serial).

\subsection{Human Participation and Evaluation}
\label{section:introduction:motivation:assistance:versus}

We may consider two main approaches
regarding human participation to a computer-based music composition process:

\begin{itemize}

\item {\em autonomous generation} --
Some recent examples are Amper, AIVA or Juke\-deck systems/companies,
based on various techniques (e.g., rules, reinforcement learning, deep learning)
aimed at the creation of original music for commercials and documentaries.
In such systems, generation is {\em automated},
with the user being restricted to a role of
{\em parametrization} of the system though a set of characteristics (style, emotion targeted, tempo, etc.).

\item {\em composition assistance} --
An example is the FlowComposer environment\footnote{Using various techniques
	such as Markov models, constraints and rules,
	and not (yet) deep learning techniques.}
\cite{papadopoulos:flow:composer:cp:2016}.
In such highly interactive systems, the user is composing (and producing) music, {\em incrementally}, with the help
(suggestion, completion, complementation, etc.) of the environment.

\end{itemize}


Most current works using deep learning to generate music are autonomous
and the way to evaluate them is a musical Turing test,
i.e. presenting to various human evaluators (beginners or experts)
original music (of a given style of a known compositor, e.g.,
Bach\footnote{The fact that Bach music is often used
	for such experiments
	may not be only because of
	its
	wide availability,
	but also because his music is actually
	easier to automate, as Bach himself was somehow an algorithmic music composer.
	An example is the way he was composing chorales by
	designing
	and applying
	(with talent)
	counterpoint rules to existing melodies.})
mixed with music generated after having learnt that style.
As we will see in the following, deep learning techniques turn out to be very efficient at succeeding in such tests,
due to their capacity to learn very well musical style from a given corpus and to generate new music that fits
this style.

As pointed out, e.g., in \cite{mgbdlcd:ncaa:2018},
most current neural networks/deep learning-based systems are black-box autonomous generators,
with low capacity for incrementality and interactivity\footnote{Two exceptions
	will be introduced in Section~\ref{section:system:deepbach}.}.
However, expert users may use them as generators of primary components (e.g., melodies, chord sequences, or/and rhythm loops)
and assemble and orchestrate them by hand.
An example is the experiment conducted by the YACHT dance music band with the
MusicVAE architecture\footnote{To be described in
	Section~\ref{section:system:musicvae}.}
from the Google Magenta project \cite{matisse:yacht:magenta:ars:tecnica:2019}.


%

\section{A First Example}
\label{section:first:example}

\label{section:system:bach:doodle}

\begin{figure}
\includegraphics[width=0.7\textwidth]{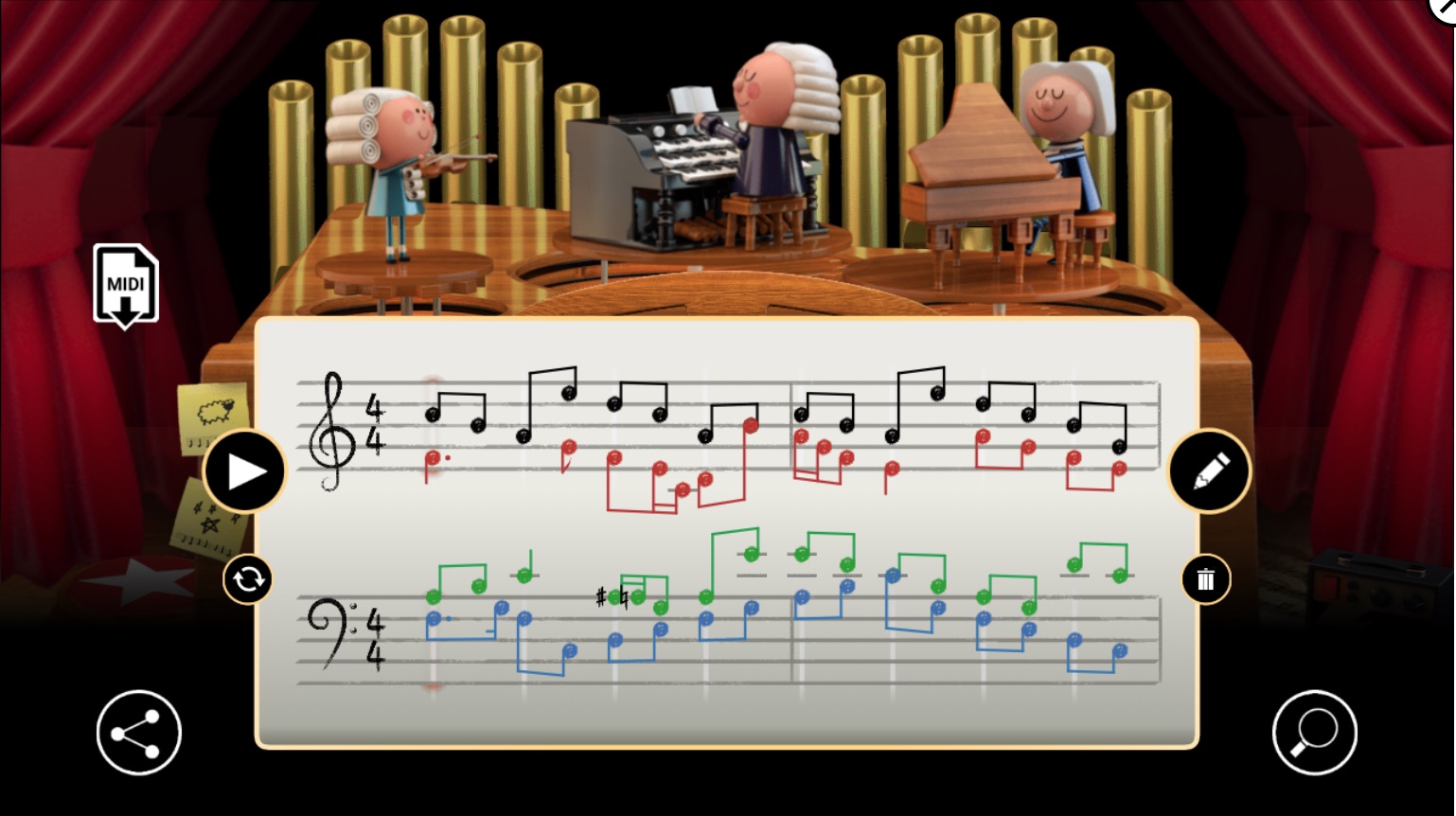}
\caption{Example of chorale generation by Bach Doodle. The original soprano melody is in black
and the generated counterpoint melodies in color (alto in red, tenor in green and bass in blue).
\copyright~2019, Google LLC, used with permission}
\label{figure:bach:doodle}
\end{figure}

On the 21st of March of 2019, for the anniversary of Bach's birthday,
Google presented an interactive Doodle
generating some Bach's style counterpoint for a melody entered interactively by the user \cite{google:bach:doodle:2019}.
In practice, the system generates three matching parts, corresponding to alto, tenor and bass voices,
as shown in Figure~\ref{figure:bach:doodle}.
The underlying architecture, named Coconet \cite{huang:counterpoint:convolution:ismir:2017},
has been trained on a dataset of 306 Bach chorales.
It will be described in Section~\ref{section:system:cononet},
but,
in this section, we will
at first
consider a more straightforward architecture, named MiniBach\footnote{MiniBach is an over simplification
	of DeepBach \cite{hadjeres:deep:bach:arxiv:2017}, to be described in Section~\ref{section:system:deepbach}.}
\cite[Section~6.2.2]{briot:dlt4mg:springer:2019}.

\label{section:system:minibach}

As a further simplification,
we consider only 4 measures long excerpts from the corpus.
Therefore, the dataset is constructed by extracting all possible 4 measures long excerpts from the original 352 chorales, also transposed in all possible keys.
%
Once trained on this dataset, the system may be used to generate three counterpoint voices corresponding to an arbitrary 4 measures long
melody provided as an input.
Somehow, it does capture the practice of
Bach, who chose various melodies for a soprano voice and composed the three additional voices melodies
(for alto, tenor and bass) in a counterpoint\index{Counterpoint} manner.

The input as well output representations are symbolic, of the piano roll type,
with a direct encoding into
one-hot vectors\footnote{{\em Piano roll format} and {\em one-hot encoding} will be
	explained in Section~\ref{section:representation}.}.
Time quantization (the value of the time step) is set at the sixteenth note,
which is the minimal note duration\index{Note!duration} used in the corpus,
i.e. there are 16 time steps for each 4/4 measure.
The resulting input representation,
which corresponds to the soprano melody,
has
the following
size: 21 possible notes $\times$ 16 time steps $\times$ 4 measures
= 1,344.
The output representation,
which corresponds to the concatenation of the three generated counterpoint melodies,
has
the following
size: (21 + 21 + 28) $\times$ 16 $\times$ 4 = 4,480.

\begin{figure}
\includegraphics[width=0.85\textwidth]{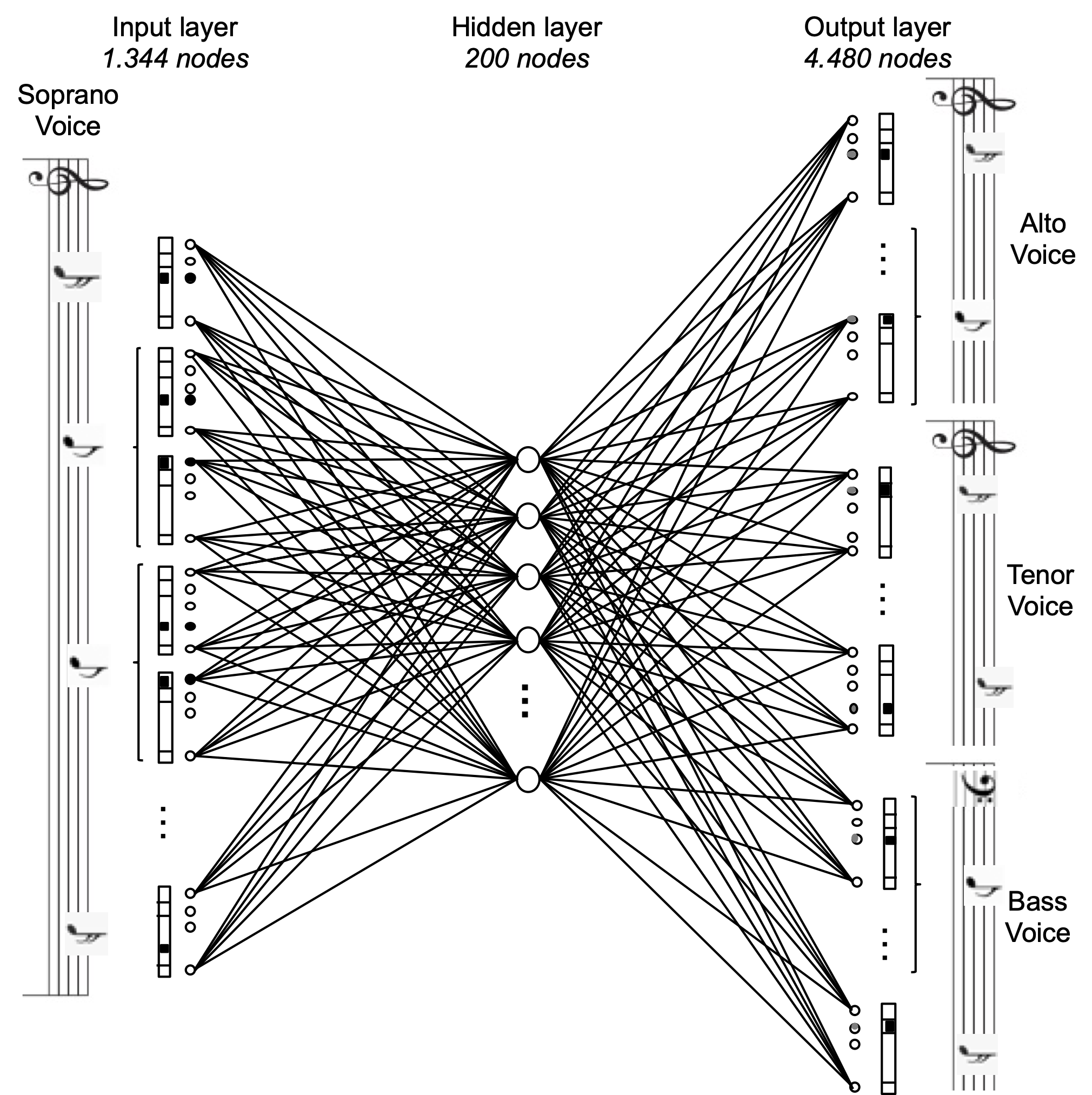}
\caption{MiniBach architecture and encoding}
\label{figure:architecture:mini:bach}
\end{figure}

The architecture,
shown in Figure~\ref{figure:architecture:mini:bach},
is feedforward
(the most basic
type of artificial neural network architecture)
for a multiple classification task:
to find out the most likely note for each time slice of the three counterpoint melodies.
There is a single hidden layer with 200 units\footnote{This is
	an arbitrary choice.}.
Successive melody time slices are encoded into successive one-hot vectors which are concatenated and directly mapped to the input nodes.
In Figure~\ref{figure:architecture:mini:bach},
each blackened vector element, as well as each corresponding blackened input node element,
illustrate the specific encoding (one-hot vector index) of a specific note time slice, depending on its actual pitch
(or a note hold in the case of a longer note, shown with a bracket). 
The dual process happens at the output.
Each grey output node element illustrates the chosen note (the one with the highest probability),
leading to a corresponding one-hot index, leading ultimately to a sequence of notes for each counterpoint voice.
(For more details, see \cite[Section~6.2.2]{briot:dlt4mg:springer:2019}.)

\begin{figure}
\includegraphics[width=0.7\textwidth]{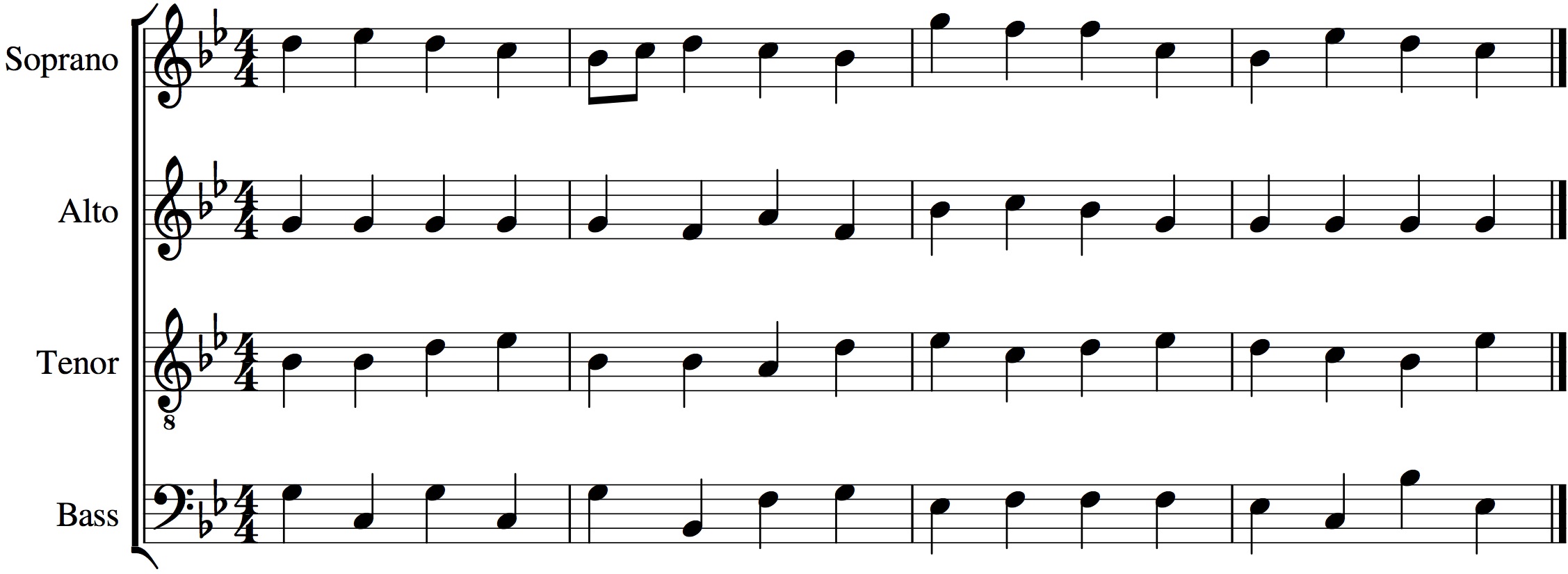}
\caption{Example of a chorale counterpoint generated
by MiniBach
from a soprano melody}
\label{figure:example:mini:bach}
\end{figure}

After training on several examples, generation can take place,
with an example of chorale counterpoint generated from a soprano melody shown in Figure~\ref{figure:example:mini:bach}.

\section{Pioneering Works and their Offsprings}
\label{section:history}

As pointed out by Pons in \cite{pons:nn:music:history:web:2018},
a first wave of applications of artificial neural networks to music
appeared in the late 1980s\footnote{A collection
	of such early papers is 
	\cite{todd:music:connectionism:book:1991}.}.
This corresponds to the second wave of the artificial neural networks movement\footnote{After
	the early stop of the first wave,
	due to the critique of the limitation (only linear classification)
	of the Perceptron \cite{book:perceptrons}.}
\cite[Section~1.2]{goodfellow:deep:learning:book:2016},
with the innovation of hidden layer(s) and backpropagation
\cite{rumelhart:pdp:1:1986,mcclelland:pdp:2:1986}.

\subsection{Todd's Time-Windowed and Conditioned Recurrent Architectures}
\label{section:system:todd}

The experiments by Todd in \cite{todd:connectionist:composition:1989}
were one of the very first attempts at exploring how to use artificial neural networks to generate music.
Although the architectures he proposed are not directly used nowadays,
his experiments and discussion were pioneering and are still an important source of information. 

\begin{figure}
\includegraphics[width=0.3\textwidth]{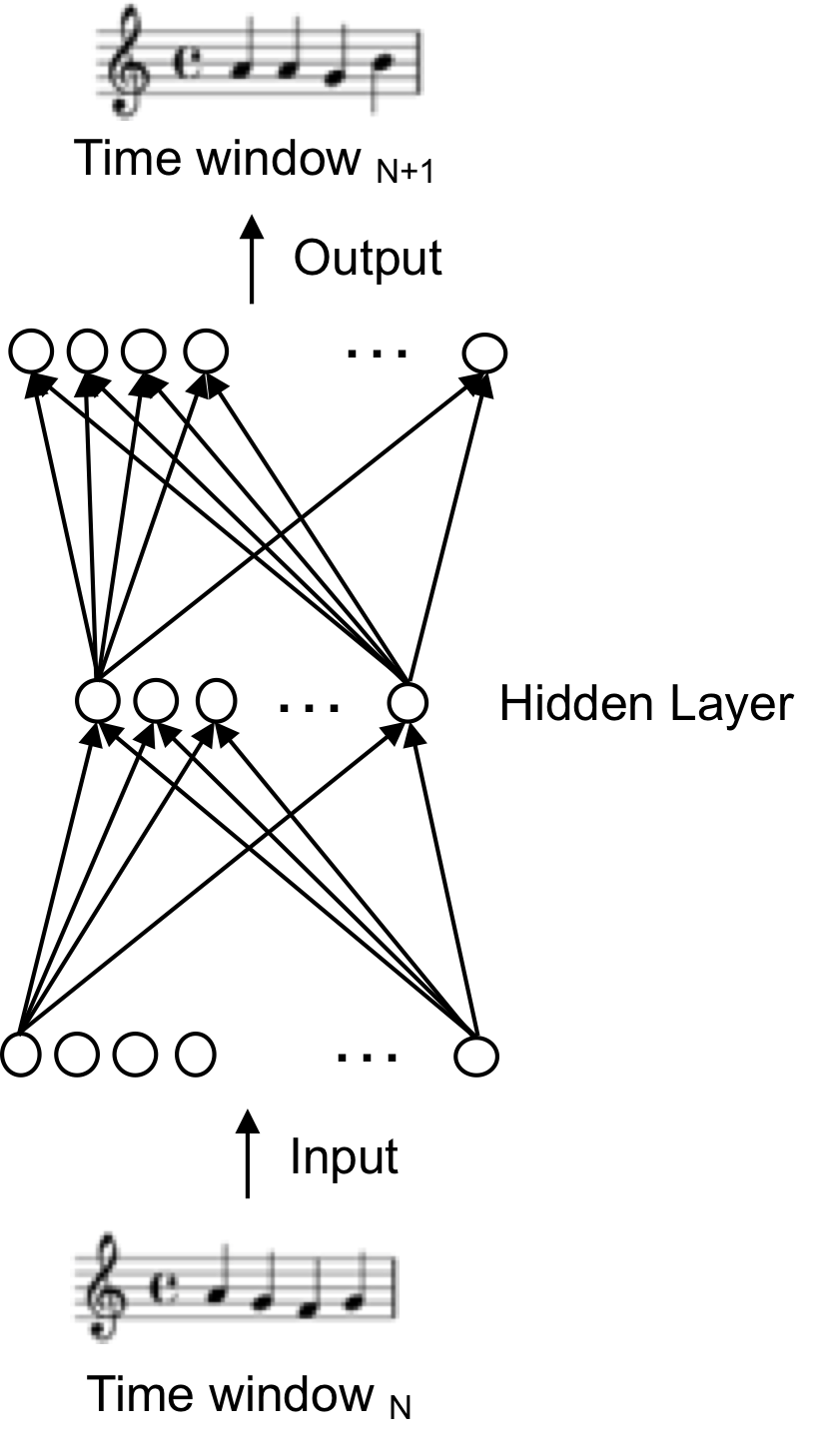}
\includegraphics[width=0.5\textwidth]{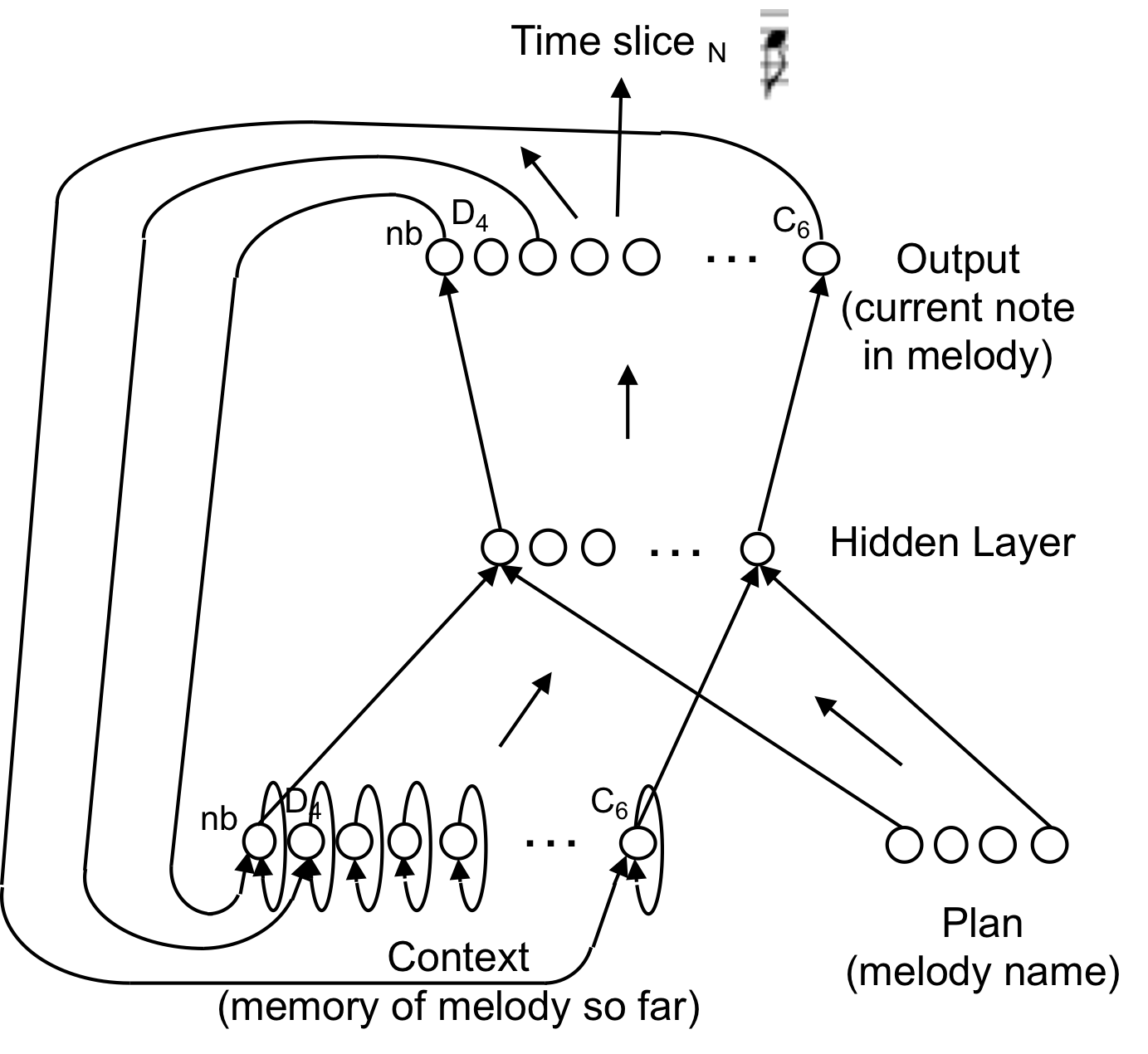}
\caption{(left) Time-Windowed architecture. (right) Sequential architecture.
Adapted
from \cite{todd:connectionist:composition:1989}}
\label{figure:todd:time:windowed:architecture}
\label{figure:todd:final:architecture}
\end{figure}

Todd's objective was to generate a monophonic melody in some iterative way.
He named his first design the Time-Windowed\index{Time-Windowed} architecture,
shown in
the left part of
Figure~\ref{figure:todd:time:windowed:architecture},
where a sliding window of successive time-periods of fixed size is considered (in practice, one measure long).
Generation is conducted iteratively melody segment by segment
(and recursively, as current output segment is entered as the next input segment and so on).
Note that, although the network will learn the pairwise correlations between
two successive melody segments\footnote{In that respect, the Time-Windowed model is analog to an order 1 Markov model
	(considering only the previous state) at the level of a melody measure.},
there is no explicit memory for learning long term correlations.


%

His third design is named Sequential\index{Sequential} and is shown in
the right part of
Figure~\ref{figure:todd:final:architecture}.
The input layer is divided in two parts, named the {\em context} and the {\em plan}.
The context is the actual memory (of the melody generated so far)
and consists in units corresponding to each note (D$_4$ to C$_6$),
plus a unit about the note begin information (notated as ``nb'' in Figure~\ref{figure:todd:final:architecture})\footnote{As a way to
	distinguish a longer note from a repeated note.}.
Therefore, it receives information from the output layer
which produces next note,
with a reentering connexion corresponding to each unit\footnote{Note that
	the output layer is isomorphic to the context layer.}.
In addition, as Todd explains it: ``A memory of more than just the single previous output (note)
is kept by having a self-feedback connection on each individual context unit.''\footnote{This is
	a peculiar characteristic of this architecture,
	as in recent standard recurrent network architecture
	recurrent connexions are encapsulated within the hidden layer (as we will see in Section~\ref{section:architecture:rnn}).
	The argument by Todd in \cite{todd:connectionist:composition:1989}
	is that context units are more interpretable than hidden units:
	``Since the hidden units typically compute some complicated,
	often uninterpretable function of their inputs,
	the memory kept in the context units will likely also be uninterpretable.
	This is in contrast to [this] design,
	where, as described earlier,
	each context unit keeps a memory of its corresponding output unit, which is interpretable.''}
The plan is a way to name\footnote{In practice,
	it is a scalar real value, e.g., 0.7,
	but Todd discusses his experiments with other possible encodings \cite{todd:connectionist:composition:1989}.}
a particular melody (among many) that the network has learnt.



Training is done by selecting a plan (melody) to be learnt.
The activations of the context units are initialized to 0
in order to begin with a clean empty context.
The network is then feedforwarded
and its output, corresponding to the first time step note, is compared to the first time step note of the melody to be learnt,
resulting in the adjustment of the weights.
The output values\footnote{Actually, as an optimization,
	Todd proposes in the following of his description to pass back the target (training) values and not the output values.}
are passed back to the current context.
And then, the network is feedforwarded again,
leading to the next time step note, again compared to the melody target,
and so on until the last time step of the melody.
This process is then repeated for various plans (melodies).

Generation of new melodies is conducted by feedforwarding the network with a new plan,
corresponding to a new melody
(not part of the training plans/melodies).
The activations of the context units are initialized to 0
in order to begin with a clean empty context.
Generation takes place iteratively, time step after time step.
Note that, as opposed to the currently more common recursive generation strategy (to be detailed in Section~\ref{section:strategy:recursive:iterative:feedforward}),
in which the output is explicitly reentered (recursively) into the input of the architecture,
in Todd's Sequential architecture the reentrance is {\em implicit} because of the specific nature of the recurrent connexions:
the output is reentered into the context units while the input
(the plan melody)
is constant.

\begin{figure}
{\large o$_A$)} \includegraphics[width=0.5\textwidth]{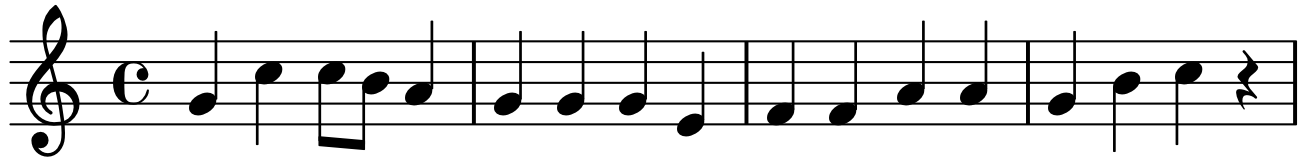}\\
{\large o$_B$)} \includegraphics[width=0.5\textwidth]{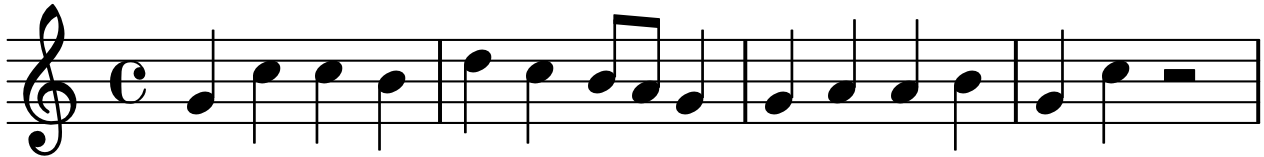}\\
{\large i$_1$)} \includegraphics[width=0.7\textwidth]{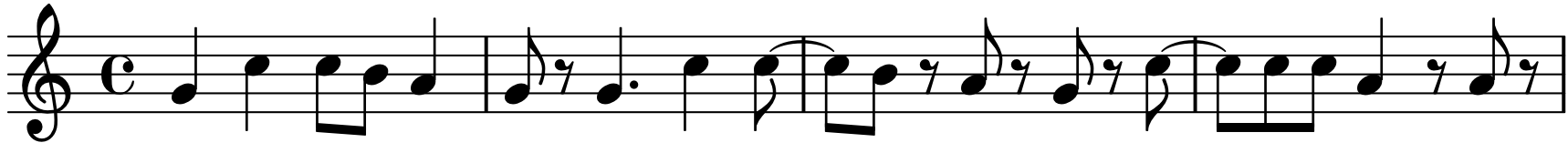}\\
\caption{Examples of melodies generated by the Sequential architecture.
(o$_A$ and o$_B$) Original plan melodies learnt.
(i$_1$)
Melody generated by interpolating between o$_A$ plan and o$_B$ plan melodies.
Adapted
from \cite{todd:connectionist:composition:1989}}
\label{figure:todd:example:interpolation}
\end{figure}

After having trained the network on a plan melody,
various melodies may be generated by extrapolation by inputing new plans,
%
or by interpolation between several (two or more) plans melodies that have been learnt.
An example of interpolation is shown in Figure~\ref{figure:todd:example:interpolation}.

\subsubsection{Influence}
\label{section:system:todd:influence}

Todd's Sequential architecture is one of the first examples of using a recurrent architecture and an iterative strategy\footnote{These
	types, as well as other types,
	of architectures and generation strategies will be
	more systematically analyzed in
	Sections~\ref{section:framework}
	and~\ref{section:basic:architecture:strategy}.}
for music generation.
Moreover, note that introducing an extra input, named plan, which represents a melody that the network has learnt,
could be seen as a precursor of {\em conditioning} architectures,
where a specific {\em additional} input is used to
{\em condition} (parametrize) the training of the architecture\footnote{An example is
	to condition the generation of a melody
	on a chord progression\index{Chord!progression},
	in the MidiNet\index{MidiNet} architecture \cite{yang:midinet:ismir:2017},
	to be described in Section~\ref{section:system:midinet}.}.
	
Furthermore,
in the
Addendum of the republication of his initial paper \cite[pages~190--194]{todd:connectionist:composition:music:connectionism:book:1991},
Todd mentions some issues and directions:

\begin{itemize}

\item {\em structure and hierarchy} -- ``One of the largest problems with this sequential network approach
	is the limited length of sequences that can be learned and the corresponding lack of global structure
	that new compositions exhibit.
	Hierarchically organized and connected sets of sequential networks
	hold promise for addressing these difficulties. (\ldots)
	One solution to these problems is first to take the sequence to be learned and divide it up
	into appropriate chunks (\ldots).''

\item {\em multiple time/clocks} -- ``Of course, one way to present this subsequence-generating network
	with the appropriate sequence of plans is to generate {\em those} by another sequential network, operating at a slower time scale.''
 
\end{itemize}

\begin{figure}
\includegraphics[width=0.7\textwidth]{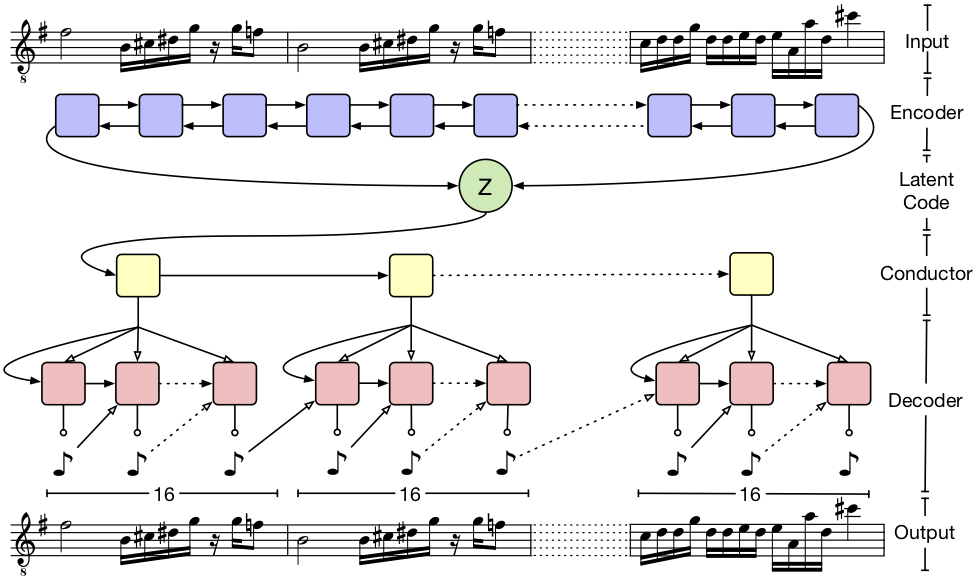}
\caption{MusicVAE architecture.
Reproduced from \cite{roberts:hierarchical:latent:icml:2018} with permission of the authors}
\label{figure:music:vae:architecture}
\end{figure}
\begin{figure}
\includegraphics[width=0.7\textwidth]{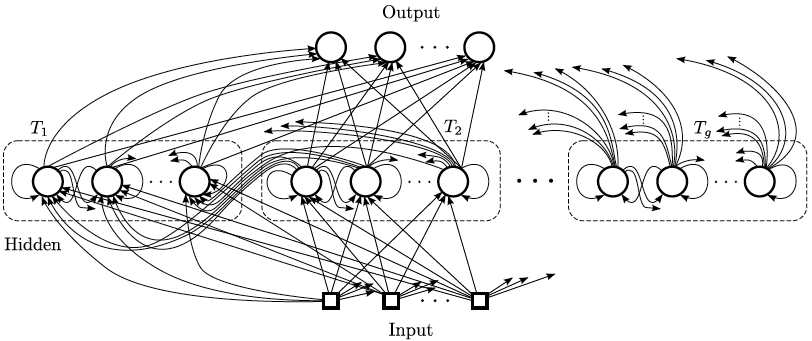}
\caption{Clockwork RNN architecture.
The RNN-like hidden layer is partitioned into several modules each with its own clock rate.
Neurons in faster module $i$ are connected
to neurons in a slower module $j$ only if a clock period $T_i < T_j$.
Reproduced from \cite{koutnik:clockworkrnn:arxiv:2014} with permission of the authors}
\label{figure:music:clockworkrnn:architecture}
\end{figure}

Thus, these early designs may be seen as precursors of some recent proposals:

\begin{itemize}

\item hierarchical architectures, such as MusicVAE \cite{roberts:hierarchical:latent:icml:2018}
	(shown in Figure~\ref{figure:music:vae:architecture} and described in Section~\ref{section:system:musicvae}); and

\item architectures with multiple time/clocks, such as
	Clockwork RNN \cite{koutnik:clockworkrnn:arxiv:2014}
	(shown in Figure~\ref{figure:music:clockworkrnn:architecture})
	and SampleRNN \cite{mehri:samplernn:arxiv:2017}.

\end{itemize}

\subsection{Lewis' Creation by Refinement}
\label{section:system:lewis}

In \cite{lewis:creation:refinement:icnn:1988},
Lewis
introduced a novel way of creating melodies, that he named {\em creation by refinement (CBR)},
by 
``reverting'' the standard way of using gradient descent for the standard task
-- adjust the connexion {\em weights} to {\em minimize} the {\em classification error} --,
into a very {\em different} task -- adjust the {\em input} in order to make the {\em classification output} turn out {\em positive}.

In his described initial experiment \cite{lewis:creation:refinement:icnn:1988},
the architecture
is a conventional feedforward neural network architecture used for binary classification, to classify ``well-formed'' melodies.
The input is a 5-note melody, each note being among the 7 notes (from C to B, without alteration).

\begin{figure}
\includegraphics[width=0.3\textwidth]{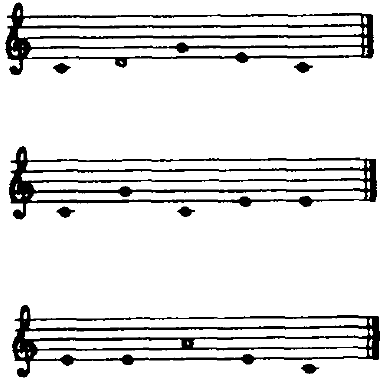}
\includegraphics[width=0.3\textwidth]{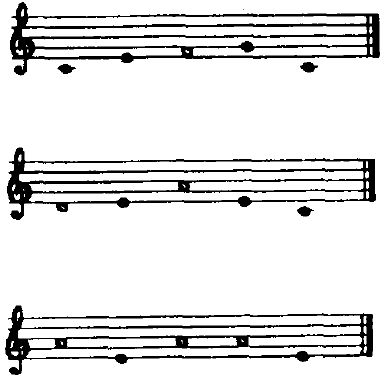}
\caption{Creation by refinement. (left) Some ``well formed'' training examples. (right) Some examples of melodies generated.
Reproduced from \cite{lewis:creation:refinement:icnn:1988} with permission of the author.
\copyright~1988, IEEE, all rights reserved}
\label{figure:cbr:training:generated:examples}
\end{figure}

For the training phase,
Lewis manually constructed 30 examples of what he meant by ``well-formed'' melodies:
using only the following intervals between notes: unison, 3rd and 5th;
and also following some scale degree stepwise motion
(some training examples are shown in the left part of the Figure~\ref{figure:cbr:training:generated:examples}).
He also constructed examples of poorly-formed melodies, not respecting the principles above.
The training phase of the network is therefore conventional,
by training it with the positive (well-formed) and negative examples that have been constructed.

\begin{figure}
\includegraphics[width=0.8\textwidth]{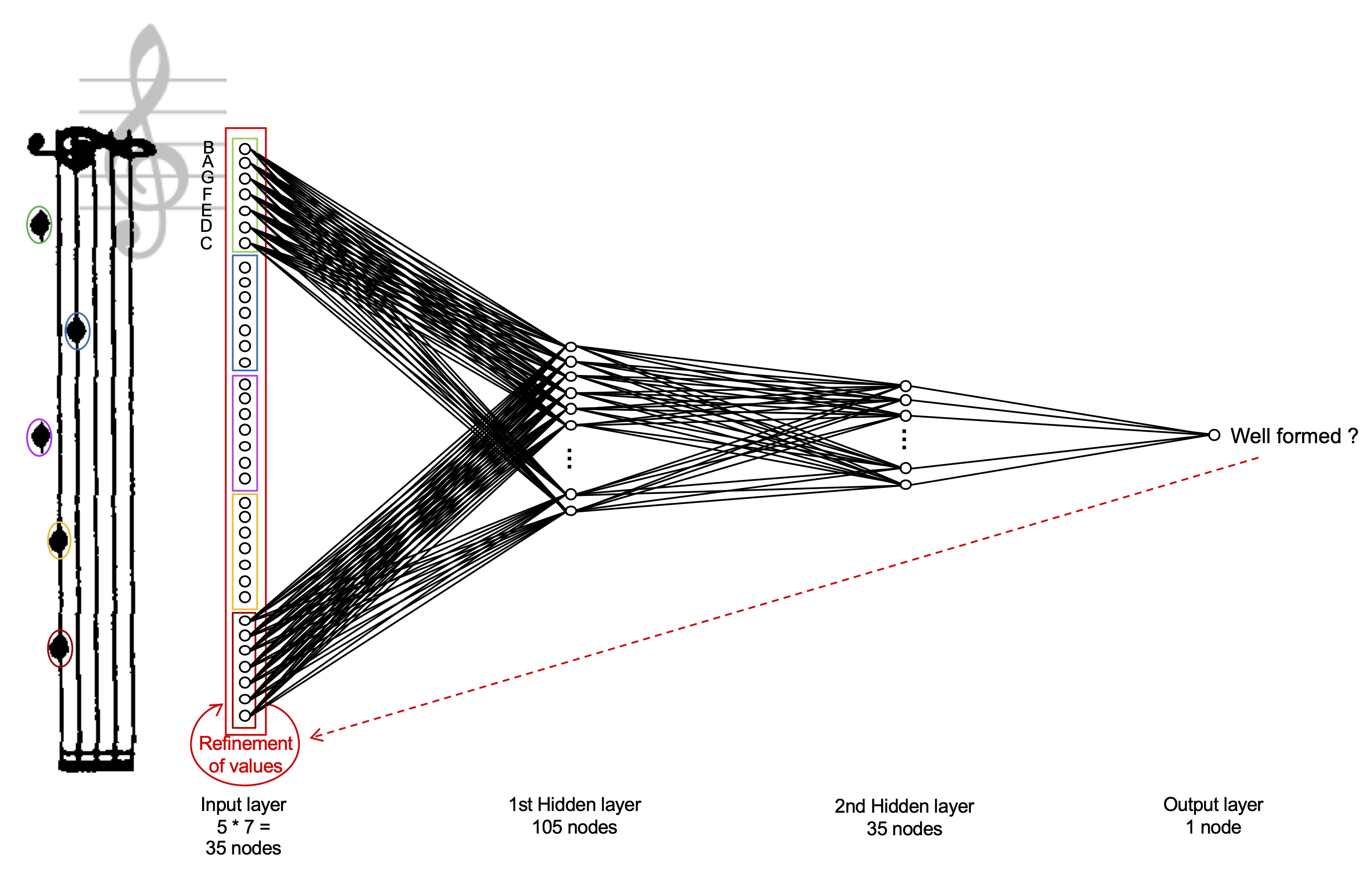}
\caption{Creation by refinement -- Architecture and strategy}
\label{figure:cbr:architecture}
\end{figure}

For the creation by refinement phase,
a vector of random values is produced, as values of the input nodes of the network.
Then, a gradient descent optimization is applied iteratively
to {\em refine} these values\footnote{Actually, in his article,
	Lewis does not detail the exact representation he uses (if he is using a one-hot encoding for each note)
	and the exact nature of refinement, i.e., adjustment of the values.}
in order to maximize a positive classification
(as shown in Figure~\ref{figure:cbr:architecture}).
This will create a new melody which is classified as well-formed.
The process may be done again, generating a new set of random values,
and controlling their adjustment in order to create a new well-formed melody.
The right part of Figure~\ref{figure:cbr:training:generated:examples} shows some examples of generated melodies.
Lewis interpretes the resulting creations as the fact that the network learned some preference for stepwise and triadic motion.

\subsubsection{Influence}
\label{section:system:lewis:influence}

The approach of creation by refinement by Lewis in 1988 can be seen as the precursor
of various approaches for controlling the creation of a content by {\em maximizing some target property}.
Examples of target properties are:

\begin{itemize}

\item maximizing a {\em positive classification} (as a well-formed melody),
in Lewis' original creation by refinement proposal \cite{lewis:creation:refinement:icnn:1988};

\item maximizing the {\em similarity\index{Similarity}} to a given {\em target\index{Target}},
in order to create a consonant\index{Consonant} melody,
as in DeepHear
\cite{sun:deep:hear};

\item maximizing\index{Maximize} the {\em activation} of a specific {\em unit},
to amplify some visual element associated to this unit,
as in Deep Dream\index{Deep Dream}
\cite{google:dream:web:2015};

\item maximizing the {\em content similarity} to some initial image {\em and} the {\em style similarity} to a reference style image,
to perform {\em style transfer\index{Style!transfer}}
\cite{gatys:neural:style:2015};

\item maximizing the {\em similarity} of the {\em structure} to some reference music,
to perform {\em style imposition\index{Style!imposition}}
\cite{lattner:structure:polyphonic:generation:jcms:2018},
as will be detailed in Section~\ref{section:systems:c-rbm}.

\end{itemize}

Interestingly, this is done by reusing
standard training mechanisms,
namely {\em backpropagation\index{Backpropagation}} to compute the gradients,
as well as {\em gradient descent\index{Gradient!descent}} (or ascent) to minimize the cost
(or to maximize the objective).


Furthermore,
in his extended article
\cite{lewis:creation:refinement:music:connectionism:book:1991},
Lewis proposed a mechanism of {\em attention} and also of {\em hierarchy}:
``In order to partition a large problem into manageable subproblems,
we need to provide both an attention mechanism to select subproblems to present to the network
and a context mechanism to tie the resulting subpatterns together into a coherent whole.'' and
``The author's experiments have employed {\em hierarchical} CBR.
In this approach, a developing pattern is recursively filled in using a scheme somewhat analogous
to a formal grammar rule such as {\em ABC $\rightarrow$ AxByC.}
which expands the string without modifying existing tokens.''

The idea of an {\em attention mechanism},
although not yet very developed,
may be seen as a precursor of attention mechanisms in
deep learning architectures.
It has initially been proposed as an {\em additional} mechanism to focus on elements of an input sequence during the training phase
\cite[Section~12.4.5.1]{goodfellow:deep:learning:book:2016},
notably for language translation applications.
More recently, it has been proposed as the {\em fundamental and unique} mechanism
(as a full alternative to recurrence or to convolution) in the Transformer architecture
\cite{vaswani:attention:transformer:arxiv:2017},
with its application to music generation,
named MusicTransformer
 \cite{huang:music:transformer:arxiv:2018}.


\subsection{From Neural Networks to Deep Learning}
\label{section:from:nn:dl}

Since
the third wave of artificial neural networks,
named {\em deep learning},
experiments on music generation
are taking advantage of:
huge {\em processing power},
optimized and standardized {\em implementations} (platforms),
and availability of {\em data},
therefore allowing experiments at large or even very large scale.
But, as we will see in Section~\ref{section:refined:architecture:strategy},
novel types of architectures have also been proposed.
Before that,
we will introduce a conceptual framework in order to help at organizing, analyzing and classifying the various types of architectures,
as well as the various usages of artificial neural networks for music generation.

\section{Conceptual Framework}
\label{section:framework}

This {\em conceptual framework} (initially proposed in \cite{briot:dlt4mg:springer:2019})
is aimed at helping the analysis of the various perspectives (and elements)
leading to the design of different deep learning-based music generation systems\footnote{{\em Systems}
	refers to the various proposals (architectures, systems and experiments)
	about deep learning-based music generation, surveyed from the literature.}.
It includes: five main {\em dimensions} (and their facets) to characterize different ways
of applying deep learning techniques to generate musical content,
and the associated {\em typologies} for each dimension.
In this paper, we will simplify the presentation and focus on the most important
aspects.

\subsection{The 5 Dimensions}
\label{section:method:dimension}


\begin{itemize}

\item {\em Objective\index{Objective}}: the nature of the musical content to be generated,
as well at its destination and use.
Examples are: melody, polyphony, accompaniment;
in the form of a musical score to be performed by some human musician(s) or an audio file to be played.

\item {\em Representation}: the nature, format and encoding of the information
(examples of music)
used to train
and to generate musical content.
Examples are: waveform signal, transformed signal (e.g., a spectrum, via a Fourier transform),
piano roll, MIDI, text;
encoded in scalar variables or/and in one-hot vectors.

\item {\em Architecture}: the nature of the assemblage of processing units
(the artificial neurons) and their connexions.
Examples are: feedforward, recurrent, autoencoder, generative adversarial networks (GAN).

\item {\em Requirement}: one of the qualities that may be desired for music generation.
Some are easier to achieve, e.g., content or length variability,
and some other ones are deeper challenges \cite{mgbdlcd:ncaa:2018}, e.g., control, creativity or structure.


\item {\em Strategy}: the way the architecture will process representations in order
to generate\footnote{It is important to highlight that,
	in this conceptual framework, by strategy we only consider the {\em generation strategy},
	i.e., the strategy to generate musical content.
	The training strategy could be quite different and is out of direct concern in this classification.}
 the objective while matching desired requirements.
Examples are: single-step feedforward, iterative feedforward,
decoder feedforward,
sampling,
creation by refinement.

\end{itemize}

Note that these five dimensions are {\em not} completely orthogonal (unrelated).
The exploration of these five different dimensions and of their interplay is actually at the core of our analysis.

%

\subsection{The Basic Generation Steps}
\label{section:generation:steps}

The basic steps for generating music, according to the {\em objective}, are as follows:

\begin{enumerate}

\item select (curate) a {\em corpus} (a set of {\em training examples}, representative of the {\em style} to be learnt);

\item select a type of {\em representation} and a type(s) of {\em encoding} and apply them to the {\em examples};

\item select a type(s) of {\em architecture} and {\em configurate} it;

\item {\em train} the {\em architecture}
with the {\em examples};


\item select a type(s) of {\em strategy} for {\em generation} and apply it to {\em generate} one or various musical contents,
and {\em decode} them into music;

\item select the {\em preferred} one(s) among the musics {\em generated}.

\end{enumerate} 

\section{Representation}
\label{section:representation}

The choice of representation and its encoding is tightly
connected to the configuration of the input and the output of the architecture, i.e. the
number of input and output nodes (variables).
Although a deep learning architecture can automatically extract
significant features from the data, the choice of representation may be significant
for the accuracy of the learning and for the quality of the generated content.

\subsection{Phases and Types of Data}
\label{section:representation:stages}

Before getting into the choices of representation
to be processed by a deep learning architecture,
it is important to identify
the main types of data to be considered, depending on the phase (training or generation):

\begin{itemize}

\item {\em training data\index{Training!data}}, the examples used as input for the training;

\item {\em generation (input) data\index{Generation!data}},
used as input for the generation
(e.g., a melody for which an accompaniment will be generated, as in the first example in Section~\ref{section:first:example}); and

\item {\em generated (output) data\index{Generated data}}, produced by the generation
(e.g., the accompaniment generated), as specified by the objective\index{Objective}.

\end{itemize}

Depending on the objective,
these two types of data may be equal or different,
e.g., in
the example in Section~\ref{section:first:example},
the generation data is a melody and the generated data is a set of (3) melodies.

\subsection{Format}
\label{section:representation:format}

The format is the nature of the representation
of a piece of music
to be interpreted by a computer.
A big divide in terms of the choice of representation
is {\em audio\index{Audio}} versus {\em symbolic\index{Symbolic}}.
This corresponds to the divide between {\em continuous\index{Continuous}} and {\em discrete\index{Discrete}} variables.
Their respective raw material is very different in nature,
as are the types of techniques for possible processing and transformation of the initial representation\footnote{In fact,
	they correspond to different scientific and technical communities,
	namely {\em signal processing\index{Signal!processing}} and {\em knowledge representation\index{Knowledge representation}}.}.
However, the actual processing of these two main types of representation
by a deep learning architecture is basically the {\em same}\footnote{Indeed, at the level of processing by a deep network architecture,
	the initial distinction between audio and symbolic representation boils down,
	as only {\em numerical} values and operations are considered.
	In this paper,
	we will focus on symbolic music representation and generation.}.

\subsubsection{Audio}
\label{section:representation:format:audio}

The main audio formats used
are:

\begin{itemize}

\item {\em signal waveform\index{Waveform}},

\item {\em spectrum}, obtained
via a {\em Fourier transform\index{Fourier transform}}\footnote{The objective of the Fourier transform
	(which could be continuous or discrete)
	is the decomposition\index{Decomposition} of an arbitrary signal\index{Signal} into its elementary components
	(sinusoidal\index{Sinusoidal} waveforms).
	As well as compressing the information, its role is fundamental for musical purposes
	as it reveals the {\em harmonic\index{Harmonics}} components of the signal.}.

\end{itemize}

The advantage of waveform is in considering the raw material untransformed, with its full initial resolution.
Architectures that process the raw signal are sometimes named {\em end-to-end\index{End-to-end architecture}} architectures.
The disadvantage is in the computational load: low level raw signal is demanding in terms of both memory and processing.
The WaveNet architecture \cite{oord:wavenet:arxiv:2016},
used for speech generation for the Google assistants,
was the first to prove the feasibility of such architectures.

\subsubsection{Symbolic}
\label{section:representation:format:symbolic}

\begin{figure}
\hspace{2.33cm}\includegraphics[width=0.4\textwidth]{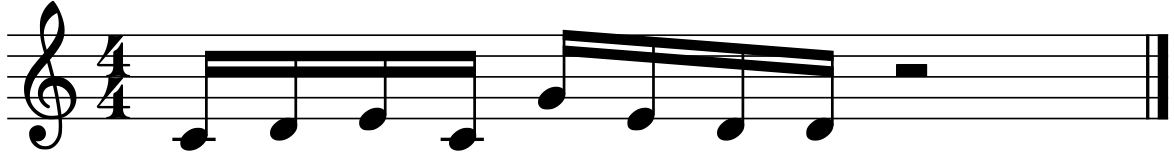}\\
\includegraphics[width=0.7\textwidth]{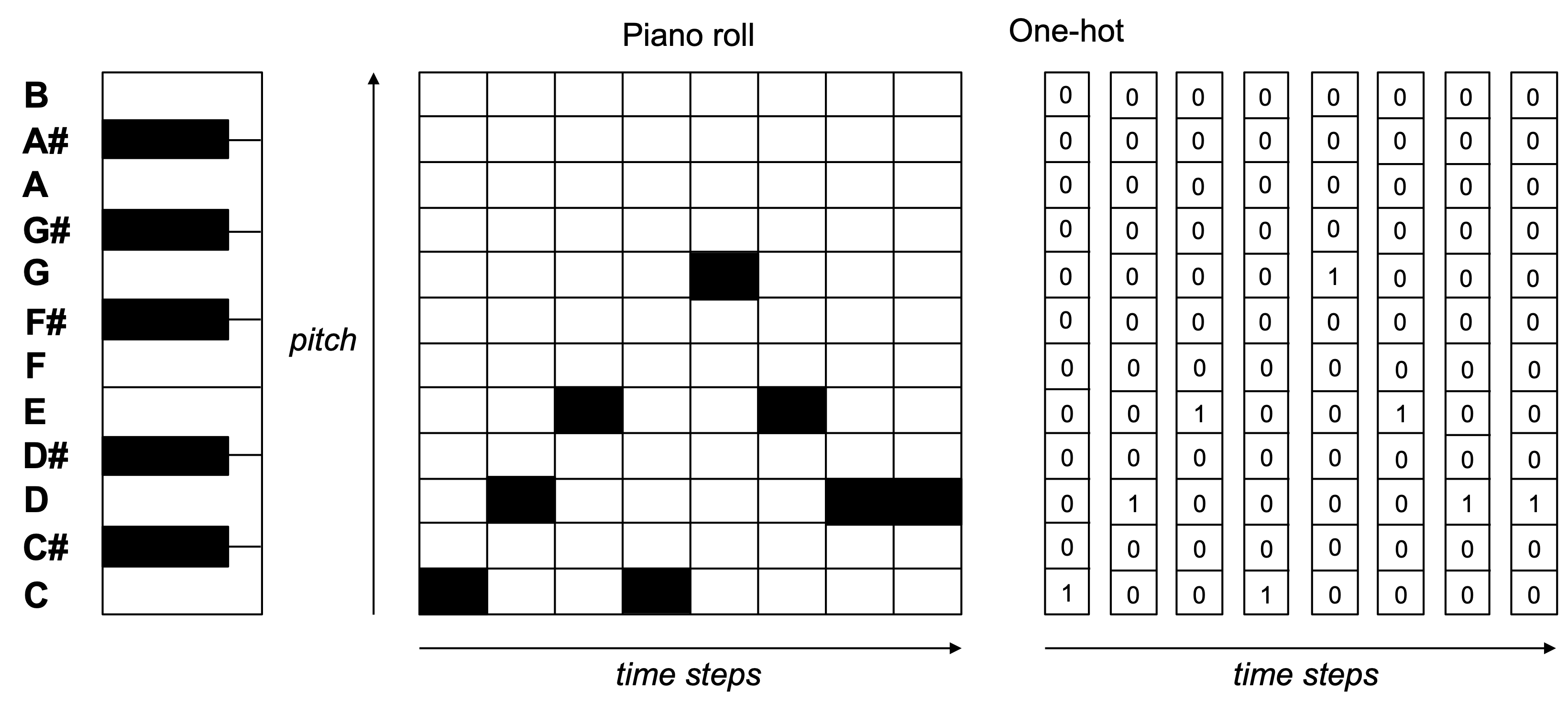}
\caption{Example of piano roll and corresponding one-hot encoding}
\label{figure:example:symbolic:piano:roll:one-hot}
\end{figure}

The main symbolic formats used are:

\begin{itemize}

\item {\em MIDI\footnote{Acronym
	of Musical Instrument Digital Interface.}} -- 
It it is a technical standard that describes a protocol based on events,
a digital interface and connectors for interoperability between various electronic musical instruments, softwares and devices
\cite{midi:web}.
Two types of MIDI event messages are considered for expressing note occurrence:
{\tt Note on} and {\tt Note off},
to indicate, respectively, the start and the end of a note played.
The MIDI {\em note number\index{MIDI!note number}},
indicates the note {\em pitch\index{Pitch}},
specified by an integer within 0 and 127.
Each note event is embedded into a data structure containing a delta-time value which also specifies the timing information,
specified as a {\em relative time} (number of periodic ticks from the beginning -- for musical scores)
or as an {\em absolute time} (in the case of {\em live performances}\footnote{The dynamics (volume)
	of a note event may also be specified.}).

\item {\em Piano roll} --
It is inspired from automated mechanical pianos
with a continuous roll of paper with perforations (holes) punched into it.
It is a two dimensional table with the x axis representing the successive time steps and the y axis the pitch,
as shown in Figure~\ref{figure:example:symbolic:piano:roll:one-hot}.

\item {\em Text} --
A significant example is the ABC notation\index{ABC notation} \cite{web:abc:notation},
a {\em de facto} standard for folk and traditional music\footnote{Note that
	the ABC notation has been
	designed {\em independently} of computer music and machine learning concerns.}.
Each note is encoded as a token, the pitch class of a note being encoded as the letter corresponding to its English notation\index{Notation convention}
(e.g., {\tt A} for A or La),
with extra notations for the octave (e.g., {\tt a'} means two octaves up)
and for the duration
(e.g., {\tt A2} means a double duration).
Measures\index{Measure} are separated by ``{\tt |}'' (bars), as in conventional scores.
An example of ABC score is shown in Figure~\ref{figure:abc:cup:tea}.

\end{itemize}

\begin{figure}
\begin{verbatim}
X: 1
T: A Cup Of Tea
R: reel
M: 4/4
L: 1/8
K: Amix
|:eA (3AAA g2 fg|eA (3AAA BGGf|eA (3AAA g2 fg|1afge d2 gf:
|2afge d2 cd|| |:eaag efgf|eaag edBd|eaag efge|afge dgfg:|
\end{verbatim}
\caption{ABC notation of ``A Cup of Tea''.
The first six lines are the header and represent {\em metadata\index{Metadata}}:
T(itle), M(eter), default note L(ength), K(ey), etc.
Reproduced from The Session \cite{web:the:session} with permission of the manager}
\label{figure:abc:cup:tea}
\end{figure}

Note that in these three cases, except for the case of
live
performances recorded in MIDI,
a {\em global time step} has to be fixed and usually corresponds,
as stated by Todd in \cite{todd:connectionist:composition:1989},
to the greatest common factor of the durations of all the notes to be
considered.

Note that each format has its pros and cons.
MIDI is probably the most versatile, as it can encompass the case of human interpretation of music (live performances),
with arbitrary/expressive timing and dynamics of note events, as, e.g., used by the Performance RNN system \cite{simon:performance:rnn:web:2017}.
But the start and the end of a long note will be represented into some very distant (temporal) positions,
thus breaking the locality of information.
The ABC notation is very compact but can only represent monophonic melodies.
In practice,
the piano roll is one of the most commonly used representations, although it has some limitations.
An important one, compared to MIDI representation, is that there is no note off information.
As a result, there is no way to distinguish between a long note and a repeated short note\footnote{Actually,
	in the original mechanical paper piano roll, the distinction is made: two holes are different from a longer single hole.
	The end of the hole is the encoding of the end of the note.}.
Main possible approaches for resolving this are:

\begin{itemize}

\item to introduce a {\em hold/replay} representation, as a dual representation of the sequence of notes
	(as used in the DeepJ system \cite{mao:deepj:arxiv:2018});

\item to divide the size of the time step\index{Time!step}
	by two
	and always mark a {\em note ending\index{Note!ending}} with a special tag
	(as used in \cite{eck:composition:lstm:2002});

\item to divide the size of the time step as
	in previous case,
	but instead
	mark a {\em new note beginning\index{Note!beginning}}
	(as used by Todd in \cite{todd:connectionist:composition:1989}, see Section~\ref{section:system:todd}); and
	
	
\item to use a special {\em hold} symbol ``\_\_'' in place of a note to specify when the previous note is held
	(as used in DeepBach \cite{hadjeres:deep:bach:arxiv:2017}, see Section~\ref{section:system:deepbach}).
	

\end{itemize}

The last solution, considering the hold symbol as a note,
is simple and uniform
but it only applies to the case of a monophonic melody.
	

\subsection{Encoding}
\label{section:representation:input:encoding}

Once the format of a representation\index{Representation} has been chosen,
the issue still remains
about
how to {\em encode} this representation.
The {\em encoding\index{Encoding}} of a representation (of a musical content)
consists in the {\em mapping} of the representation
(composed of a set of {\em variables}, e.g., pitch or dynamics)
into a set of {\em inputs} (also named {\em input nodes\index{Input!node}} or {\em input variables\index{Input!variable}})
for the neural network architecture.

There are two basic approaches:

\begin{itemize}

\item {\em value-encoding} -- A continuous, discrete or boolean variable is directly encoded as a {\em scalar}; and

\item {\em one-hot-encoding} -- A discrete or a categorical variable is encoded as a {\em categorical variable}
	through a vector with the number of all possible elements as its length.
	Then, to represent a given element, the corresponding element of the {\em one-hot vector}\footnote{The name
		comes from digital circuits,
		{\em one-hot} referring to a group of bits among which the only legal (possible) combinations of values are those
		with a single {\em high} (hot!) (1) bit, all the others being {\em low} (0).}
	is set to 1 and all other elements to 0.

\end{itemize}

For instance,
the pitch of a note could be represented as a {\em real number} (its frequency in Hertz),
an {\em integer number} (its MIDI note number),
or a {\em one-hot vector},
as shown in the right part of
Figure~\ref{figure:example:symbolic:piano:roll:one-hot}\footnote{The Figure
	also illustrates that a piano roll could be straightforwardly encoded as a sequence of one-hot vectors
	to construct the input representation of an architecture,
	as, e.g., has been shown in Figure~\ref{figure:architecture:mini:bach}.}.
The advantage of value encoding\index{Value encoding} is its compact representation,
at the cost of sensibility because of numerical operations (approximations).
The advantage of one-hot encoding\index{One-hot encoding}
(actually the most common strategy)
is its robustness\index{Robustness} against numerical operations approximations
(discrete versus analog),
at the cost of a high cardinality and therefore a potentially large number of nodes for the architecture.


\section{Main Basic Architectures and Strategies}
\label{section:basic:architecture:strategy}

For reasons of space limitation, we will now jointly introduce architectures and strategies\footnote{As a reminder
	from Section~\ref{section:method:dimension},
	we only consider here {\em generation} strategies.}.
For an alternative analysis guided by requirements (challenges), please see \cite{mgbdlcd:ncaa:2018}.

\subsection{Feedforward Architecture}
\label{section:architecture:feedforward}

The {\em feedforward architecture}\footnote{Also
	named {\em multilayer
	Perceptron}
	(MLP).}
is the most basic
and
common
type
of artificial neural
network architecture.

\subsubsection{Feedforward Strategy}

An example of use
has been detailed in Section~\ref{section:first:example}.
The generation strategy used in this example is
the most basic type of strategy,
as it consists in {\em feedforwarding} within a single step the input data into the input layer,
through successive hidden layers,
until the output layer. Therefore we name it the {\em single-step feedforward strategy},
abbreviated as {\em feedforward strategy}\footnote{The feedforward architecture and the feedforward strategy
	are naturally associated,
	although, as we will see in some of the next sections,
	other associations are possible.}.



\subsubsection{Iterative Strategy}
\label{section:strategy:iterative:feedforward}

In Todd's Time-Windowed architecture in Section~\ref{section:system:todd},
generation is processed iteratively
by feedforwarding current melody segment in order to obtain next one, and so on.
Therefore we name it the {\em iterative feedforward strategy},
abbreviated as {\em iterative strategy}.

\subsection{Recurrent Architecture}
\label{section:architecture:rnn}

\begin{figure}
\includegraphics[width=0.45\textwidth]{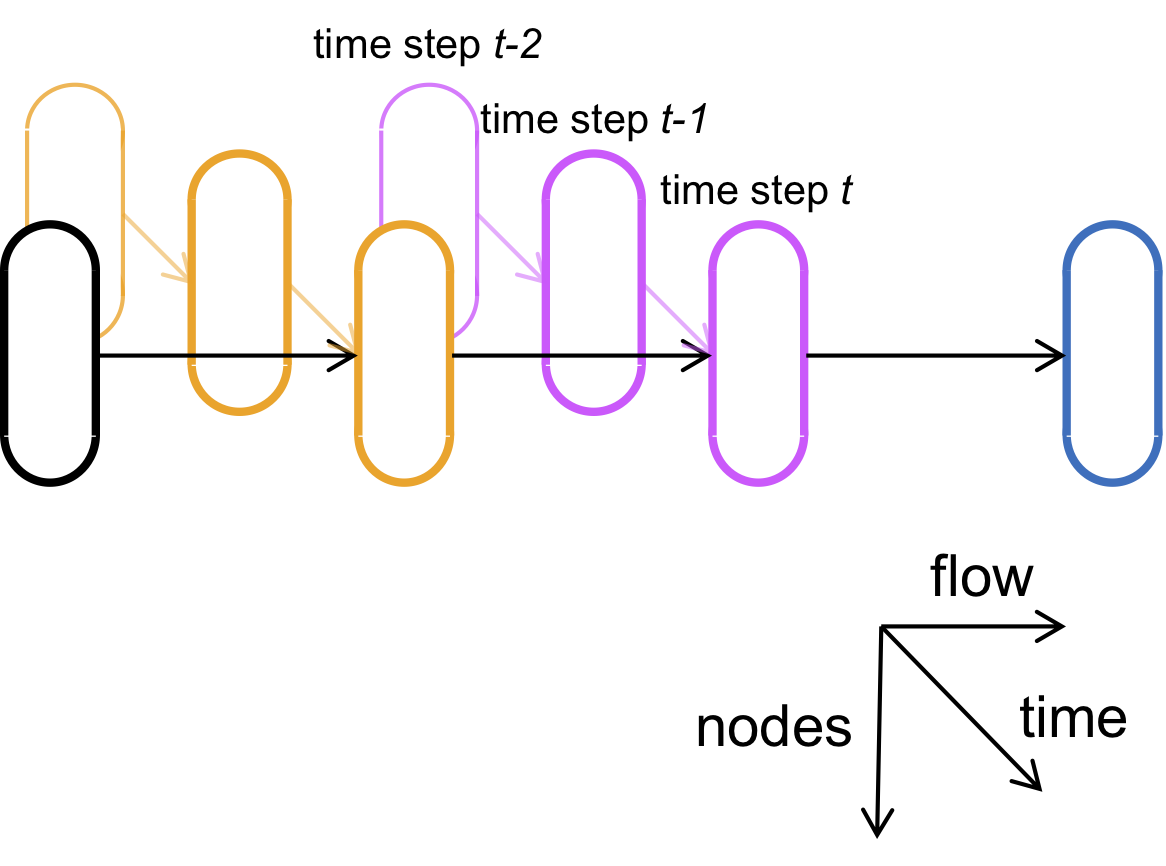}
\caption{Recurrent neural network.
Each successive layer (along the flow of computation) is represented as an oblong design (hiding the detail of its nodes).
The diagonal axis represents the time dimension,
with the previous step value of each layer in thinner and lighter color}
\label{figure:recurrent:network}
\end{figure}

A {\em recurrent neural network\index{Recurrent!neural network}} (RNN\index{RNN}) is a feedforward neural network
extended with {\em recurrent connexions\index{Recurrent!connexion}} in order to learn series of items
(e.g., a melody as a sequence of notes).
Todd's Sequential\index{Sequential} architecture
in Section~\ref{section:system:todd} is an example although not of a common type.
As pointed out in Section~\ref{section:system:todd},
in modern recurrent architectures, recurrent connexions are encapsulated within
a (hidden) layer,
which allows an arbitrary number of recurrent layers
(as shown in Figure~\ref{figure:recurrent:network}).

\subsubsection{Recursive Strategy}
\label{section:strategy:recursive:iterative:feedforward}

\begin{figure}
\includegraphics[width=0.8\textwidth]{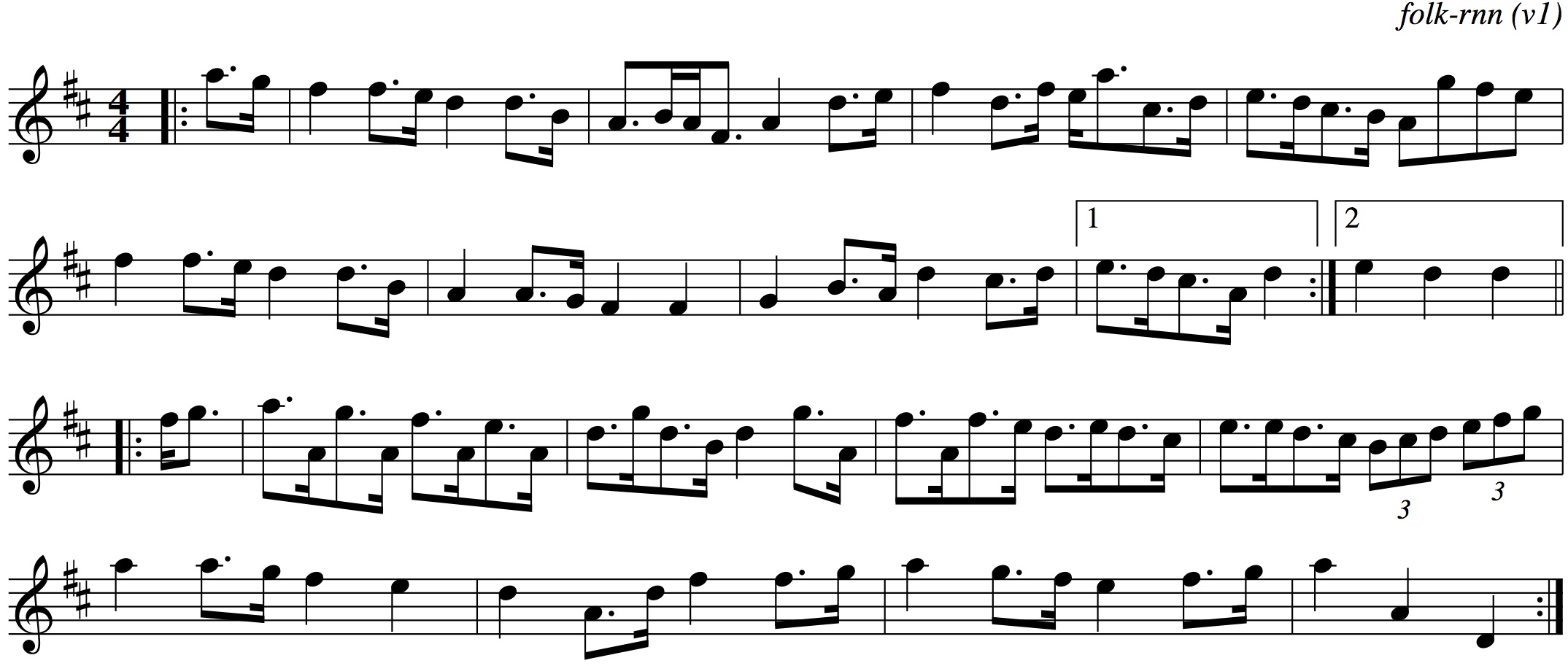}
\caption{Score of ``The Mal's Copporim'' automatically generated.
Reproduced from \cite{sturm:celtic:melody:csmc:2016} with permission of the authors}
\label{figure:score:mal:copporim}
\end{figure}

The first music generation experiment using
current state of the art of recurrent architectures,
the LSTM (Long Short-Term Memory \cite{hochreiter:lstm:1997}) architecture,
is the generation of blues chord (and melody) sequences
by Eck and
Schmidhuber
in \cite{eck:composition:lstm:2002}.
Another interesting example is the architecture by Sturm {\em et al.} to generate Celtic\index{Celtic} melodies \cite{sturm:celtic:melody:csmc:2016}.
It is trained on examples selected from the folk music repository named The Session \cite{web:the:session}
and uses text (the
ABC notation \cite{web:abc:notation}, see Section~\ref{section:representation:format:symbolic})
as the representation format.
Generation
(an example is shown in Figure~\ref{figure:score:mal:copporim})
is done using a {\em recursive
strategy},
a special case of iterative strategy,
for generating a sequence of notes (or/and chords),
as initially described for text generation by Graves in \cite{graves:generating:sequences:rnn:arxiv:2014}:


\begin{itemize}

\item select some {\em seed\index{Seed}} information as the {\em first} item (e.g., the first note of a melody);

\item {\em feedforward} it into the recurrent network in order to produce the {\em next} item (e.g., next note);

\item use this next item as the next input to produce the {\em next next} item; and

\item repeat this process iteratively until a {\em sequence} (e.g., of notes, i.e. a melody) of the desired length is produced\footnote{Note that,
	as opposed to
	feedforward strategy
	(and decoder feedforward strategy,
	to be introduced in Section~\ref{section:strategy:decoder:feedforward}),
	iterative and recursive strategies allow the generation of musical content of {\em arbitrary length}.}.

\end{itemize}

%

\subsubsection{Sample Strategy}
\label{section:strategy:iterative:sampling:feedforward}

A limitation of
applying straightforwardly the iterative feedforward strategy on a recurrent network
is that generation is {\em deterministic}\footnote{Indeed, most artificial neural networks are deterministic.
	There are stochastic versions of artificial neural networks
	-- the Restricted Boltzmann Machine (RBM) \cite{hinton:rbm:science:2006}
	is an example --
	but they are not mainstream.
	An example of use of RBM will be described in Section~\ref{section:systems:c-rbm}.}.
As a consequence, feedforwarding the {\em same input} will always produce the {\em same output}.
Indeed, as the generation of the next note, the next next note, etc., is deterministic,
the {\em same} seed note will lead to the {\em same} generated series of notes\footnote{The actual length of the melody generated
	depending on the number of iterations.}.
Moreover, as there are only 12 possible input values (the 12 pitch classes, disregarding the possible octaves),
there are only 12 possible melodies.

\begin{figure}
\includegraphics[scale=0.6]{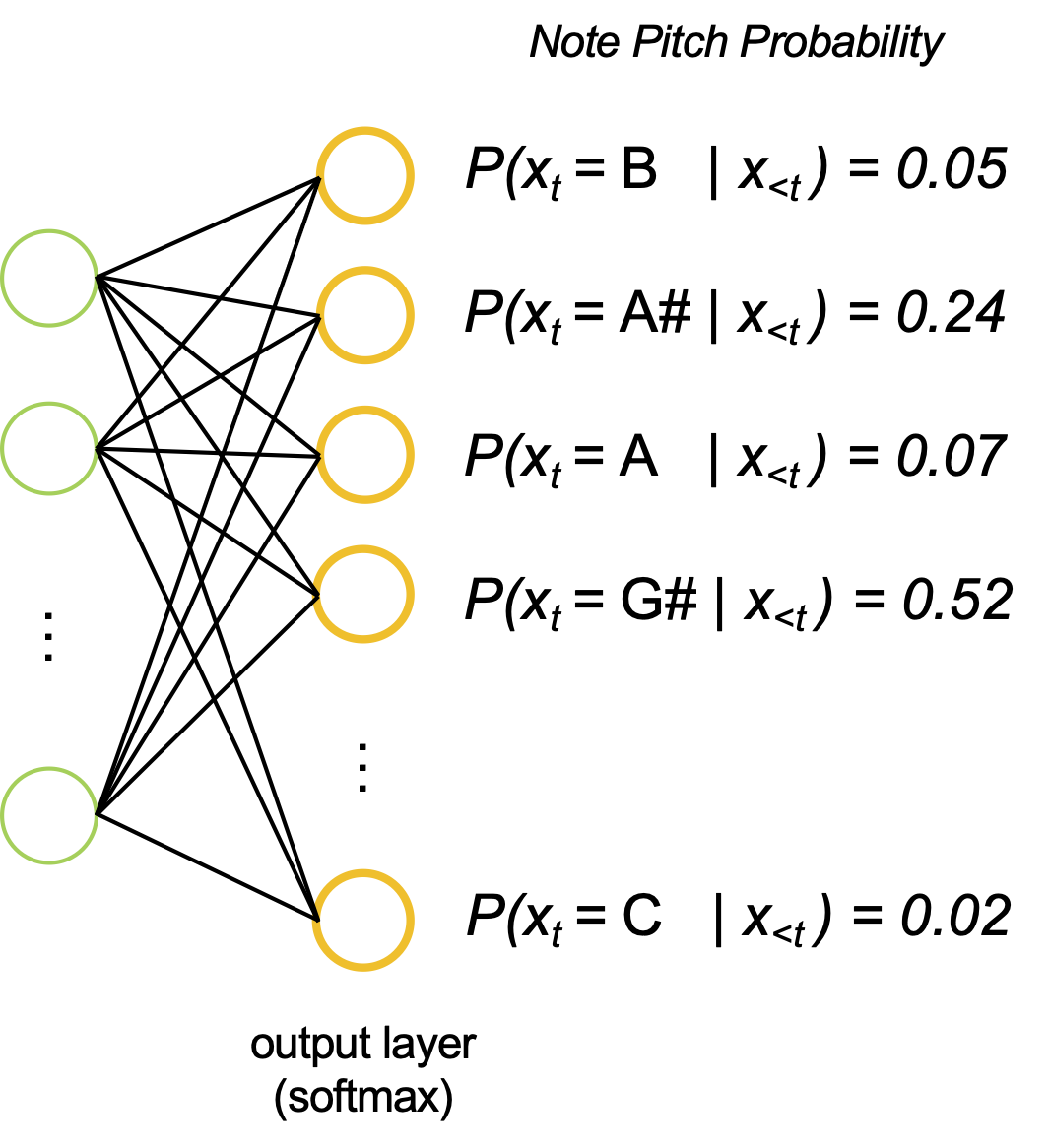}
\caption{A softmax at the output layer computes the probability for each pitch}
\label{figure:softmax:output}
\end{figure}

Fortunately,
the solution is quite simple.
The assumption is that the generation is modeled as a classification\index{Classification!task} task,
i.e., the output representation of the melody is one-hot encoded
and the output layer activation function is softmax.
See an example in Figure~\ref{figure:softmax:output},
where $P(\text{x}_t = \text{C} | \text{x}_{<t})$
represents the conditional probability for the element (pitch of the note) x$_t$ at step $t$ to be a C
given the previous elements x$_{<t}$ (the melody generated so far).
The default {\em deterministic} strategy consists in choosing the pitch with the highest probability,
i.e. $\text{argmax}\index{Argmax}_{\text{x}_t} P(\text{x}_t | \text{x}_{<t})$
(that is G$\sharp$ in
Figure~\ref{figure:softmax:output}).
We can then easily switch to a {\em nondeterministic} strategy,
by {\em sampling}\footnote{Sampling
	is the action of generating an element (a {\em sample\index{Sample}})
	from a {\em stochastic\index{Stochastic}} model
	according to a {\em probability distribution\index{Probability!distribution}}.}
the output which corresponds (through the softmax function)
to a probability distribution between possible pitches.
By sampling a pitch following the distribution generated recursively by the architecture\footnote{The chance of sampling a given pitch
	is its corresponding probability. In the example shown in Figure~\ref{figure:softmax:output},
	G$\sharp$ has around one chance in two of being selected and A$\sharp$ one chance in four.},
we introduce stochasticity in the process of generation and thus {\em content variability} in the generation.



\section{Compound Architectures}
\label{section:architecture:compound}
\label{section:architecture:compound:composition:types}

For more sophisticated objectives and requirements,
{\em compound\index{Compound architecture}} architectures may be used.
%
%
We will see that, from an architectural\index{Architectural} point of view,
various types of
combination\footnote{We are taking inspiration
	from concepts and terminology in programming languages and software architectures \cite{shaw:software:architecture:book:1996},
	such as {\em refinement}, {\em instantiation}, {\em nesting} and {\em pattern} \cite{gamma:design:patterns:book:1994}.}
may be used:

\subsection{Composition}
\label{section:architecture:compound:composition}

Several architectures, of the same type or of different types, are
composed,
e.g.:

\begin{itemize}


\item a bidirectional\index{Bidirectional} RNN\index{Bidirectional!RNN|see{Bidirectional recurrent neural network}},
composing
two RNNs, forward and backward in time,
e.g.,
as used in
the C-RNN-GAN\index{C-RNN-GAN}
\cite{mogren:c-rnn-gan:arxiv:2016}
(see Figure~\ref{figure:c:rnn:gan:architecture})
and
the MusicVAE\index{MusicVAE}
\cite{roberts:hierarchical:latent:icml:2018}
(see Figure~\ref{figure:music:vae:architecture}
and Section~\ref{section:system:musicvae})
architectures;
and

\item the RNN-RBM\index{RNN-RBM} architecture \cite{boulanger:temporal:dependencies:icml:2012},
composing
an RNN\index{RNN} architecture and an RBM\index{RBM} architecture.

\end{itemize}

\subsection{Refinement}
\label{section:architecture:compound:refinement}

One architecture is refined and specialized\index{Specialization}
through some additional constraint(s),
e.g.:

\begin{itemize}


\item an autoencoder architecture (to be introduced in Section~\ref{section:architecture:autoencoder}),
which is a feedforward architecture with one hidden layer
with the same cardinality (number of nodes) for the input layer and the output layer; and

\item a variational autoencoder\index{Variational!autoencoder} (VAE\index{VAE}) architecture,
which is an autoencoder with an additional constraint on the distribution of the variables of the hidden layer
(see Section~\ref{section:architecture:vae}),
e.g., the GLSR-VAE architecture \cite{hadjeres:glsr:vae:arxiv:2017}.

\end{itemize}

\subsection{Nesting}
\label{section:architecture:compound:nested}

An architecture is nested\index{Nested} into the other one, e.g.:

\begin{itemize}

\item a stacked autoencoder\index{Stacked autoencoder} architecture\footnote{A {\em stacked autoencoder}
	is a hierarchical nesting of autoencoders with decreasing number of hidden layer units,
	as shown in the right part of
	Figure~\ref{figure:autoencoder}.},
e.g., the DeepHear architecture
\cite{sun:deep:hear};
and

\item a recurrent autoencoder architecture (Section~\ref{section:architecture:vrae}),
where an RNN\index{RNN} architecture
is nested within an autoencoder\footnote{More precisely,
	an RNN is nested within the encoder\index{Encoder}
	and another RNN within the decoder\index{Decoder}.
	Therefore, it is also named an RNN Encoder-Decoder\index{RNN Encoder-Decoder} architecture.},
e.g., the MusicVAE\index{MusicVAE} architecture \cite{roberts:hierarchical:latent:icml:2018}
(see Section~\ref{section:system:musicvae}).

\end{itemize}

%
%
%


\subsection{Pattern}
\label{section:architecture:compound:pattern}

\begin{figure}
\includegraphics[width=0.5\textwidth]{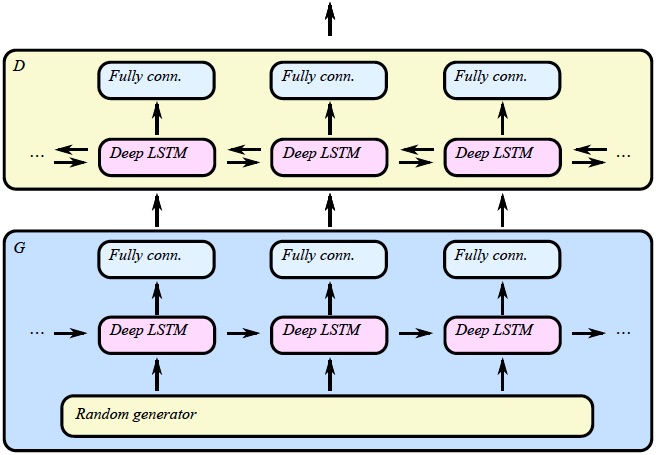}
\caption{C-RNN-GAN architecture with the D(iscriminator) GAN component being a bidirectional RNN (LSTM).
Reproduced from \cite{mogren:c-rnn-gan:arxiv:2016} with permission of the authors}
\label{figure:c:rnn:gan:architecture}
\end{figure}

An architectural pattern\index{Architectural!pattern} is instantiated\index{Instantiate} onto a given architecture(s)\footnote{Note
	that we limit here the scope of a pattern to the {\em external enfolding} of an existing architecture.
	Additionally, we could have considered convolutional, autoencoder and even recurrent architectures
	as an {\em internal} architectural pattern.}, e.g.:

\begin{itemize}


\item the Anticipation-RNN\index{Anticipation-RNN} architecture
\cite{hadjeres:anticipation:rnn:arxiv:2017},
that instantiates the {\em conditioning} pattern\footnote{Such as introduced by Todd
	in his Sequential architecture conditioned by a plan
	in Section~\ref{section:system:todd}.}
onto an RNN
with the output of another RNN as the conditioning input\index{Conditioning!input};
and

\item the C-RNN-GAN\index{C-RNN-GAN} architecture
\cite{mogren:c-rnn-gan:arxiv:2016},
where the {\em GAN\index{GAN} (Generative Adversarial Networks)}
pattern (to be introduced in Section~\ref{section:architecture:gan})
is instantiated onto two RNN\index{RNN} architectures,
the second one (discriminator) being bidirectional
(see Figure~\ref{figure:c:rnn:gan:architecture}); and

\item the MidiNet architecture \cite{yang:midinet:ismir:2017}
(see Section~\ref{section:system:midinet}),
where the {\em GAN\index{GAN}}
pattern
is instantiated onto two convolutional\footnote{Convolutional architectures
	are actually an important
	component of the current success of deep learning
	and they recently emerged as an alternative,
	more efficient to train,
	to recurrent architectures \cite[Section~8.2]{briot:dlt4mg:springer:2019}.
	A {\em convolutional} architecture is composed of a succession of feature maps and pooling layers
	\cite[Section~9]{goodfellow:deep:learning:book:2016}\cite[Section~5.9]{briot:dlt4mg:springer:2019}.
	(We could have considered convolutional as an internal architectural pattern, as has just been remarked in a previous footnote.)
	However, we do not detail convolutional architectures here, because of space limitation
	and of non specificity regarding music generation applications.}
feedforward architectures,
on which a {\em conditional} pattern
is instantiated.


\end{itemize}


\subsection{Illustration}
\label{section:architecture:compound:classification:examples}

\begin{figure}
\includegraphics[width=0.9\textwidth]{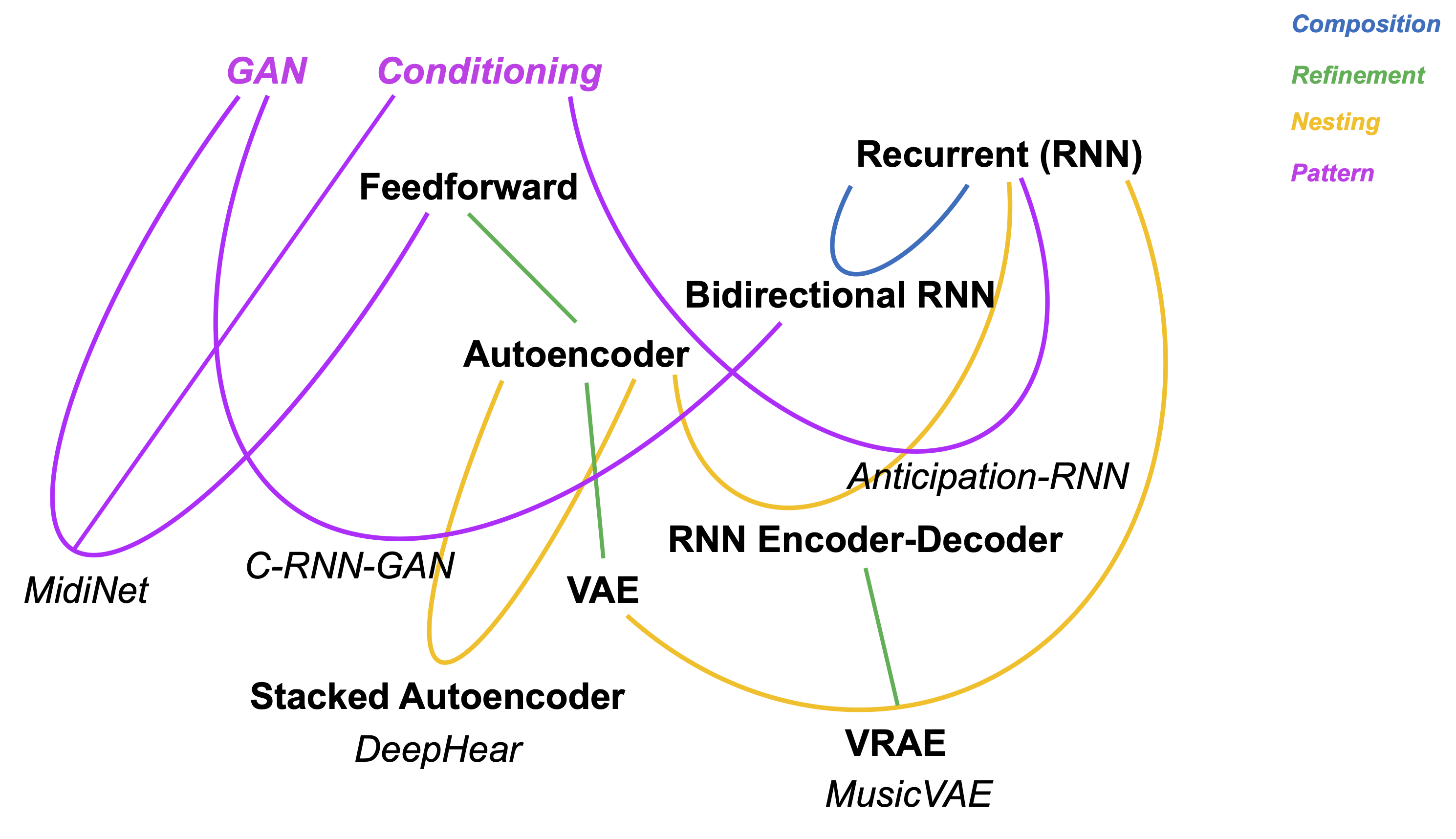}
\caption{A tentative illustration of various examples and combination types (in color fonts) of compound architectures (in black bold font)
and systems (in black italics font)}
\label{figure:architectures:classification}
\end{figure}

Figure~\ref{figure:architectures:classification} illustrates various examples of compound architectures
and of actual music generation systems.


\subsection{Combined Strategies}
\label{section:combined:strategies}

Note that the strategies for generation can be combined too,
although not in the same way as the architectures:
they are actually
used simultaneously on different components of the architecture.
In the examples discussed in Section~\ref{section:strategy:iterative:sampling:feedforward},
the {\em recursive strategy} is used by recursively feedforwarding current note into the architecture
in order to produce next note and so on,
while the {\em sampling strategy} is used at the output of the architecture to sample the actual note (pitch)
from the possible notes with their respective probabilities.


\section{Examples of Refined Architectures and Strategies}
\label{section:refined:architecture:strategy}

\subsection{Autoencoder Architecture}
\label{section:architecture:autoencoder}

\begin{figure}
\includegraphics[width=0.3\textwidth]{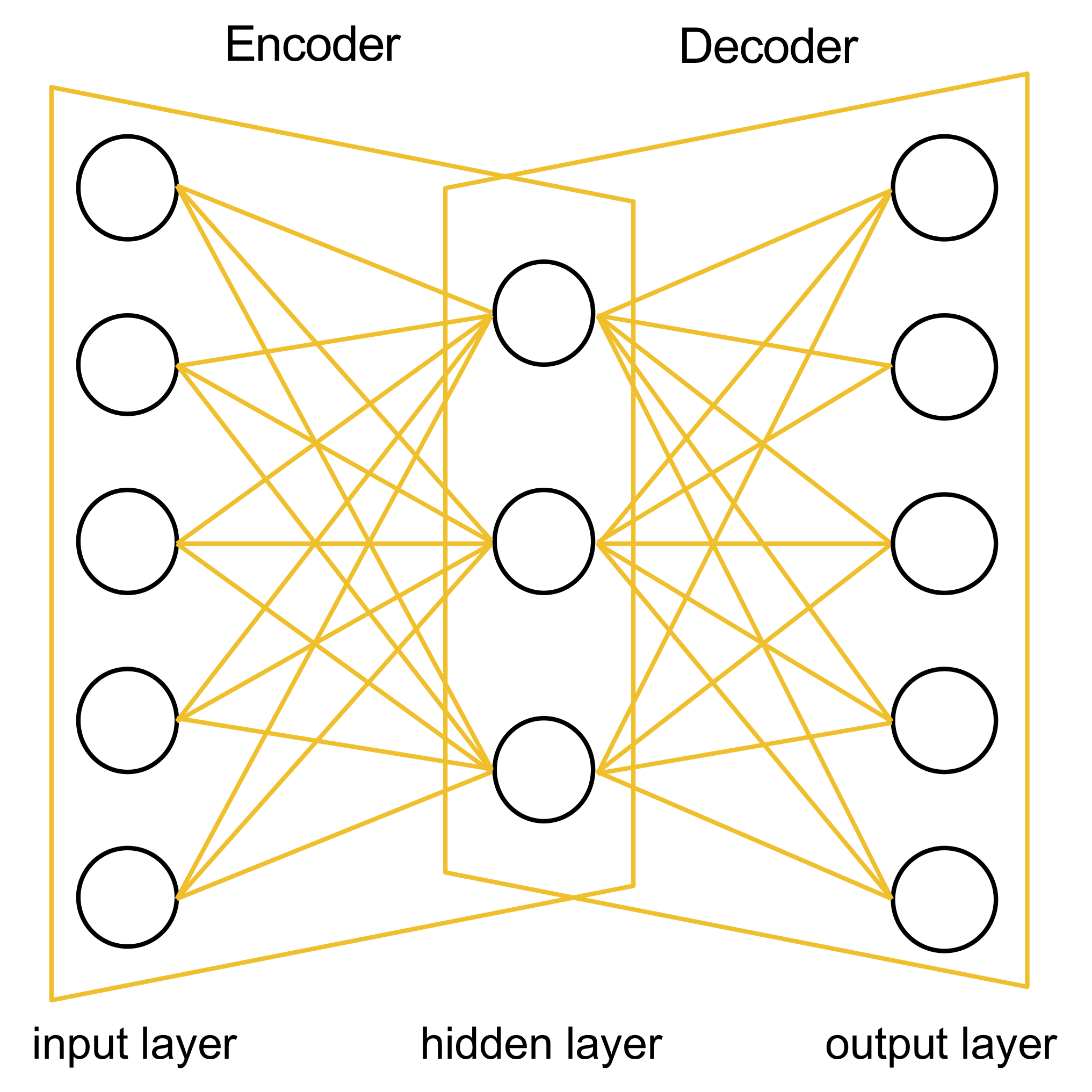}
\includegraphics[width=0.492\textwidth]{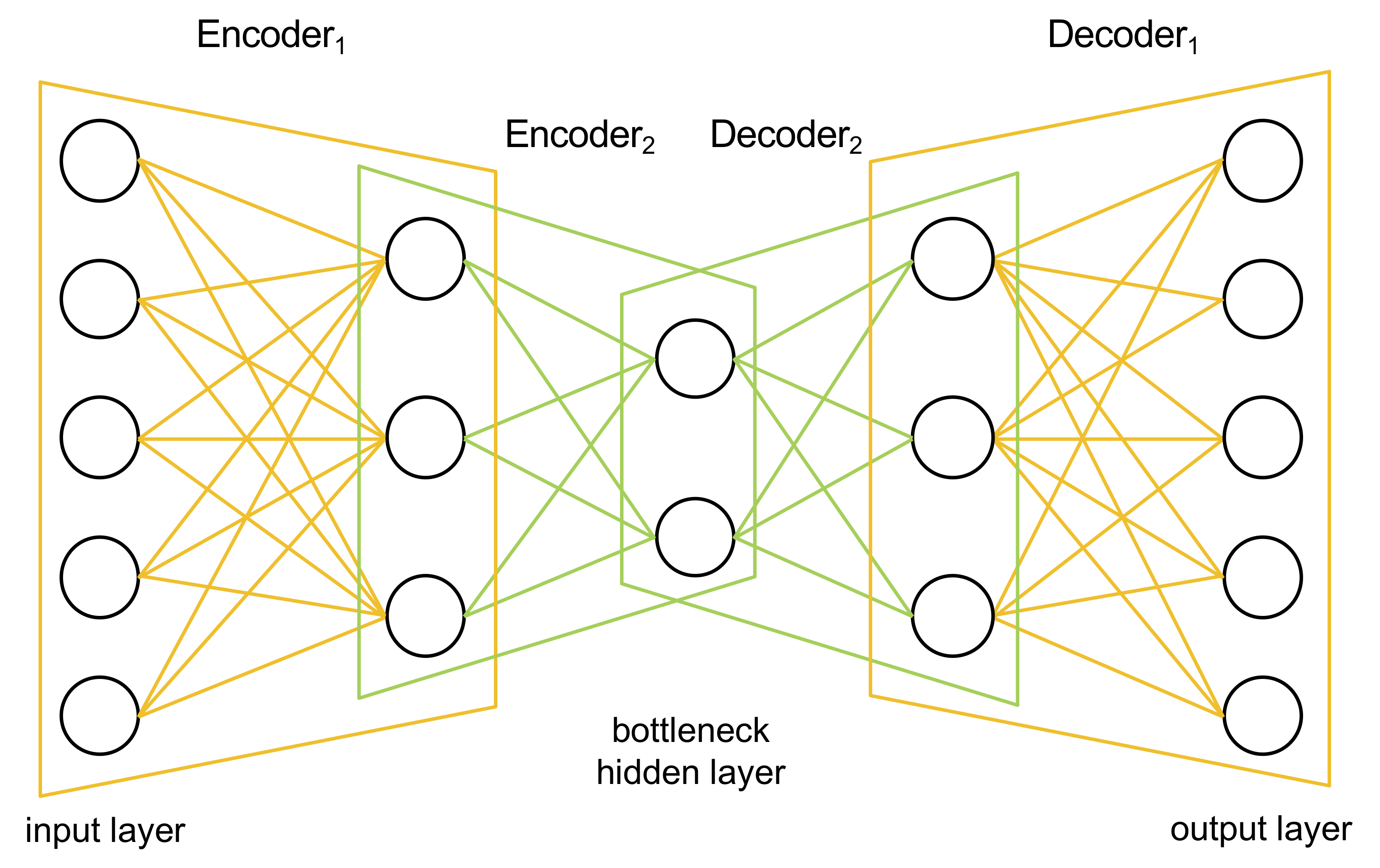}
\caption{(left) Autoencoder architecture. (right) Stacked autoencoder (order-2) architecture}
\label{figure:autoencoder}
\end{figure}

An {\em autoencoder} is a refinement of a feedforward neural network
with two constraints: (exactly) one hidden layer and
the number of output nodes\index{Output!node} is equal to
the number of input nodes\index{Input!node}.
The output layer actually {\em mirrors} the input layer,
creating its peculiar symmetric\index{Symmetric} diabolo (or sand-timer) shape aspect,
as shown in the left part of Figure~\ref{figure:autoencoder}.

An autoencoder is trained with each of the examples as the input and as the output.
Thus, the autoencoder tries to learn the identity\index{Identity} function.
As the hidden layer usually has fewer nodes than the input layer,
the {\em encoder\index{Encoder}} component
must {\em compress} information\footnote{Compared to traditional dimension reduction algorithms,
	such as principal component analysis\index{Principal component analysis} (PCA\index{PCA}),
	feature extraction is nonlinear,
	but it does not ensure orthogonality of the dimensions, as we will see in Section~\ref{section:architecture:autoencoder:disentanglement}.},
while the {\em decoder\index{Decoder}}
has to {\em reconstruct}, as accurately as possible, the initial information.
This forces the autoencoder to {\em discover} significant (discriminating) {\em features\index{Feature!extraction}} to encode\index{Encoding} useful information
into the hidden layer nodes (also named the {\em latent variables\index{Latent!variable}}).

\subsubsection{Decoder Feedforward Strategy}
\label{section:strategy:decoder:feedforward}

\begin{figure}
\includegraphics[width=0.8\textwidth]{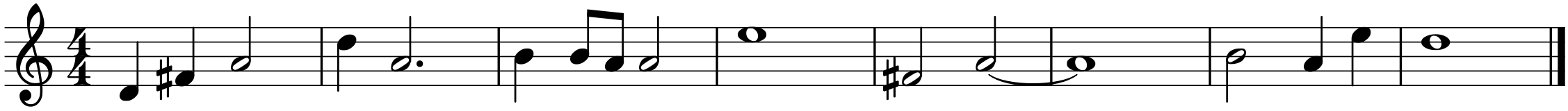}
\caption{Example of melody generated by an autoencoder trained on a Celtic melodies corpus}
\label{figure:example:ae:celtic:melody}
\end{figure}

The latent variables\index{Latent!variable} of an autoencoder constitute
a compact representation
of the common features of the learnt examples.
By instantiating these latent variables and decoding them (by feedforwarding them into the decoder), we can generate a new musical content
corresponding to the values of the latent variables, in the same format as the training examples.
We name this strategy the {\em decoder feedforward\index{Decoder!feedforward strategy} strategy}.
An example generated after training an autoencoder on a set of Celtic melodies
(selected from the
folk music repository The Session \cite{web:the:session}, introduced in Section~\ref{section:strategy:recursive:iterative:feedforward})
is shown in Figure~\ref{figure:example:ae:celtic:melody}
(see \cite{briot:compress:create:musmat:2020} for more details).
An early example of this strategy is the use of the DeepHear nested (stacked) autoencoder architecture to generate ragtime music
according to the style learnt \cite{sun:deep:hear}.

\subsection{Variational Autoencoder Architecture}
\label{section:architecture:vae}


Although producing interesting results, an autoencoder suffers from some discontinuity
in the generation when exploring the latent space\footnote{See more details,
	e.g., in \cite{rocca:understanding:vae:blog:towards:data:science:2019}.}.
A {\em variational autoencoder\index{Variational!autoencoder}} (VAE\index{VAE}) \cite{kingma:vae:arxiv:2014}
is a refinement
of an autoencoder
where, instead of encoding an example as a single point,
a variational autoencoder encodes it as a probability {\em distribution} over the latent space\footnote{The implementation
	of the encoder of a VAE
	actually generates a mean vector and a standard deviation vector \cite{kingma:vae:arxiv:2014}.},
from which the latent variables are sampled,
and with the constraint that the distribution follows some {\em prior probability distribution}\footnote{This constraint is implemented
	by adding a specific term to the cost\index{Cost} function
	to compute the cross-entropy\index{Cross-entropy} between the distribution
	of latent variables and the prior distribution.},
usually a Gaussian distribution\index{Gaussian distribution}\index{Probability!distribution}.
This regularization ensures two main properties:
{\em continuity} (two close points in the latent space should not give two completely different contents once decoded)
and {\em completeness} (for a chosen distribution, a point sampled from the latent space should provide a ``meaningful'' content once decoded)
\cite{rocca:understanding:vae:blog:towards:data:science:2019}.
The price to pay is some larger reconstruction error,
but the tradeoff between reconstruction and regularity can be adjusted
depending on the priorities.

As with an autoencoder\index{Autoencoder}, a VAE\index{VAE} will learn the identity\index{Identity} function,
but furthermore the decoder will learn the relation between the prior (Gaussian) distribution of the latent variables\index{Latent!variable}
and the learnt examples.
A very interesting characteristic
for generation purposes
is therefore in the {\em meaningful exploration} of the latent space\index{Latent!space},
as a variational autoencoder is able to learn a ``smooth'' {\em mapping}
from the latent space to realistic examples \cite{wild:disentangled:vae:web:2018}.


%
%
%
%

%
%
%
%

\subsubsection{Variational Generation}
\label{section:architecture:vae:generation}

\begin{figure}
\includegraphics[width=0.9\textwidth]{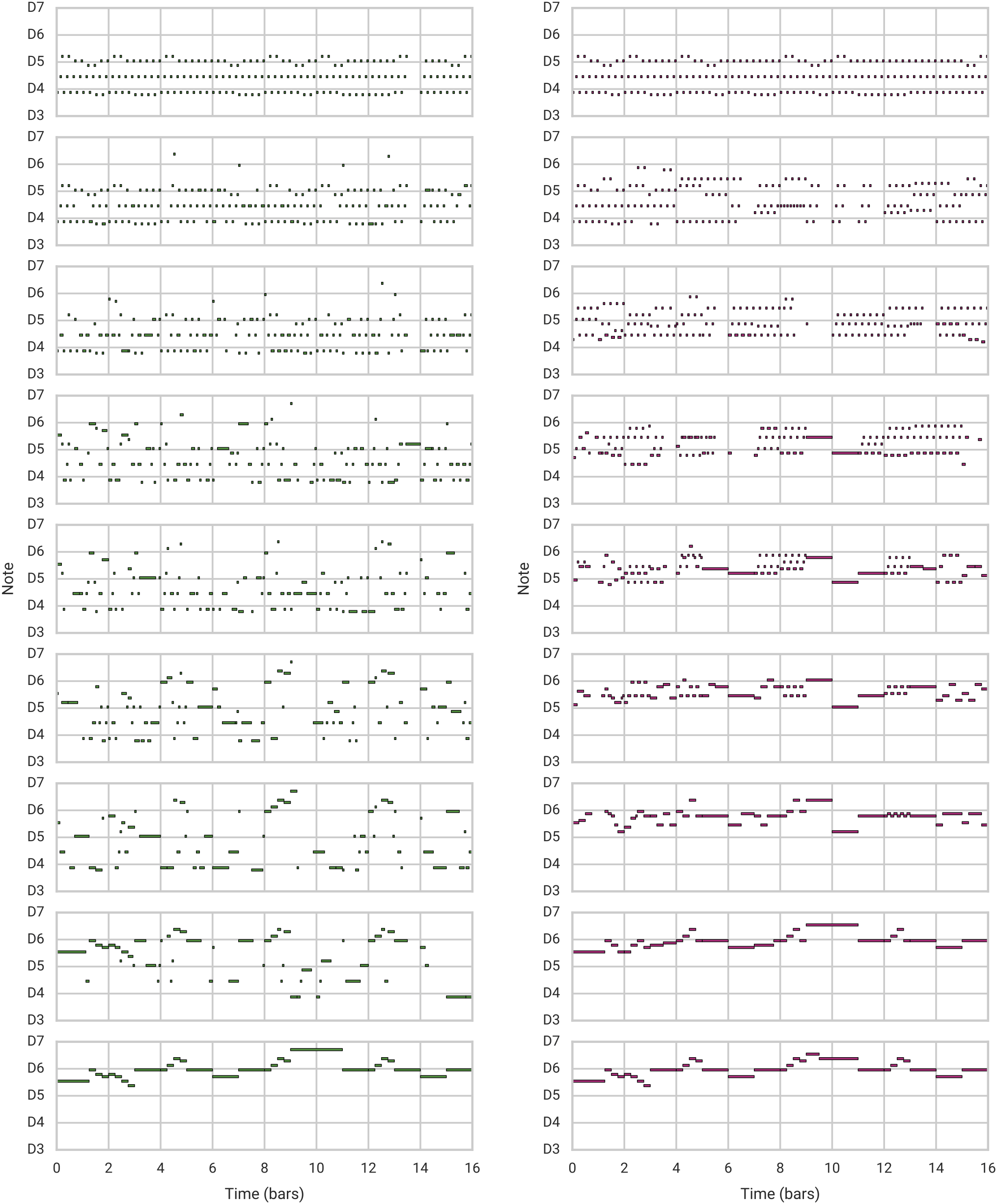}
\caption{Comparison of interpolations between the top and the bottom melodies by
(left) interpolating in the data (melody) space
and
(right) interpolating in the latent space and decoding it into melodies.
Reproduced from \cite{roberts:hierarchical:latent:arxiv:2018} with permission of the authors}
\label{figure:music:vae:example:interpolation}
\end{figure}

\begin{figure}
\includegraphics[width=0.6\textwidth]{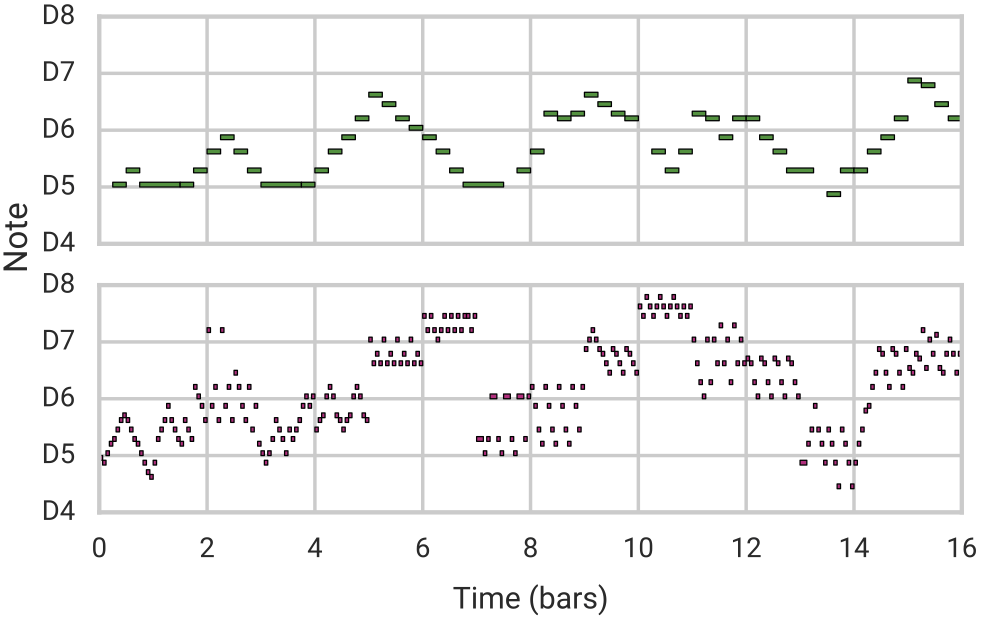}
\caption{Example of a melody generated (bottom) by MusicVAE
by adding a ``high note density'' attribute vector to the latent space of an existing melody (top).
Reproduced from \cite{roberts:hierarchical:latent:arxiv:2018} with permission of the authors}
\label{figure:music:vae:example:density}
\end{figure}

Examples of possible dimensions captured by latent variables learnt by the VAE
are
the note duration range (the distance between shortest and longest note)
and the note pitch range (the distance between lowest and highest pitch).
This latent representation (vector of latent variables)
can be used to explore the latent space\index{Latent!space}
with various operations
to control/vary the generation of content.
Some examples of operations on the latent space
(as summarized in \cite{roberts:hierarchical:latent:icml:2018})
are:

\begin{itemize}

\item {\em translation};

\item {\em interpolation\index{Interpolation}}\footnote{The interpolation in the latent space
	produces more meaningful and interesting melodies
	than the interpolation in the data space (which basically just varies the ratio of notes from the two melodies)
	\cite{roberts:hierarchical:latent:arxiv:2018},
	as shown in Figure~\ref{figure:music:vae:example:interpolation}.};

\item {\em averaging\index{Average}}; and

\item {\em attribute vector arithmetics\index{Attribute!vector arithmetics}},
	by addition or subtraction of an attribute vector capturing a given characteristic\footnote{This attribute vector is computed as
	the average latent vector for a collection of examples sharing that attribute\index{Attribute} (characteristic),
	e.g., high density of notes (see an example in Figure~\ref{figure:music:vae:example:density}),
	rapid change, high register, etc.}.

\end{itemize}

\subsubsection{Disentanglement}
\label{section:architecture:autoencoder:disentanglement}

One limitation of using a variational autoencoder is that the dimensions (captured by latent variables) are not independent ({\em orthogonal}),
as in the case of Principal component analysis (PCA).
However, various techniques have being recently proposed to improve the {\em disentanglement} of the dimensions
(see, e.g., \cite{mathieu:disentangling:disentanglement:arxiv:2019}).

Another issue is that the semantics (meaning) of the dimensions captured by the latent variables
is automatically ``chosen'' by the VAE architecture in function of the training examples and the configuration
and thus can only be interpreted {\em a posteriori}.
However, some recent approaches propose to ``force'' the meaning of latent variables,
by splitting the decoder into various components and training them onto a specific dimension (e.g., rhythm or pitch melody)
\cite{yang:analogy:disentanglement:arxiv:2019}.

%

\subsection{Variational Recurrent Autoencoder (VRAE) Architecture}
\label{section:architecture:vrae}


An interesting example of nested architecture (see Section~\ref{section:architecture:compound:nested})
is a
variational recurrent autoencoder (VRAE).
The motivation is to combine:

\begin{itemize}

\item the {\em variational} property of the VAE architecture for controlling the generation; and

\item the {\em arbitrary length} property
of the RNN architecture used with the recursive strategy\footnote{As pointed out in Section~\ref{section:strategy:recursive:iterative:feedforward},
	the generation of next note from current note
	is repeated until a sequence of the desired length is produced.}.

\end{itemize}

\label{section:system:musicvae}

An example (also hierarchical) is the MusicVAE architecture \cite{roberts:hierarchical:latent:icml:2018}
(shown in Figure~\ref{figure:music:vae:architecture},
with two examples of controlled generation in Figures~\ref{figure:music:vae:example:interpolation}
and~\ref{figure:music:vae:example:density}).
	
\subsection{Generative Adversarial Networks (GAN) Architecture}
\label{section:architecture:gan}

\begin{figure}
\includegraphics[width=0.6\textwidth]{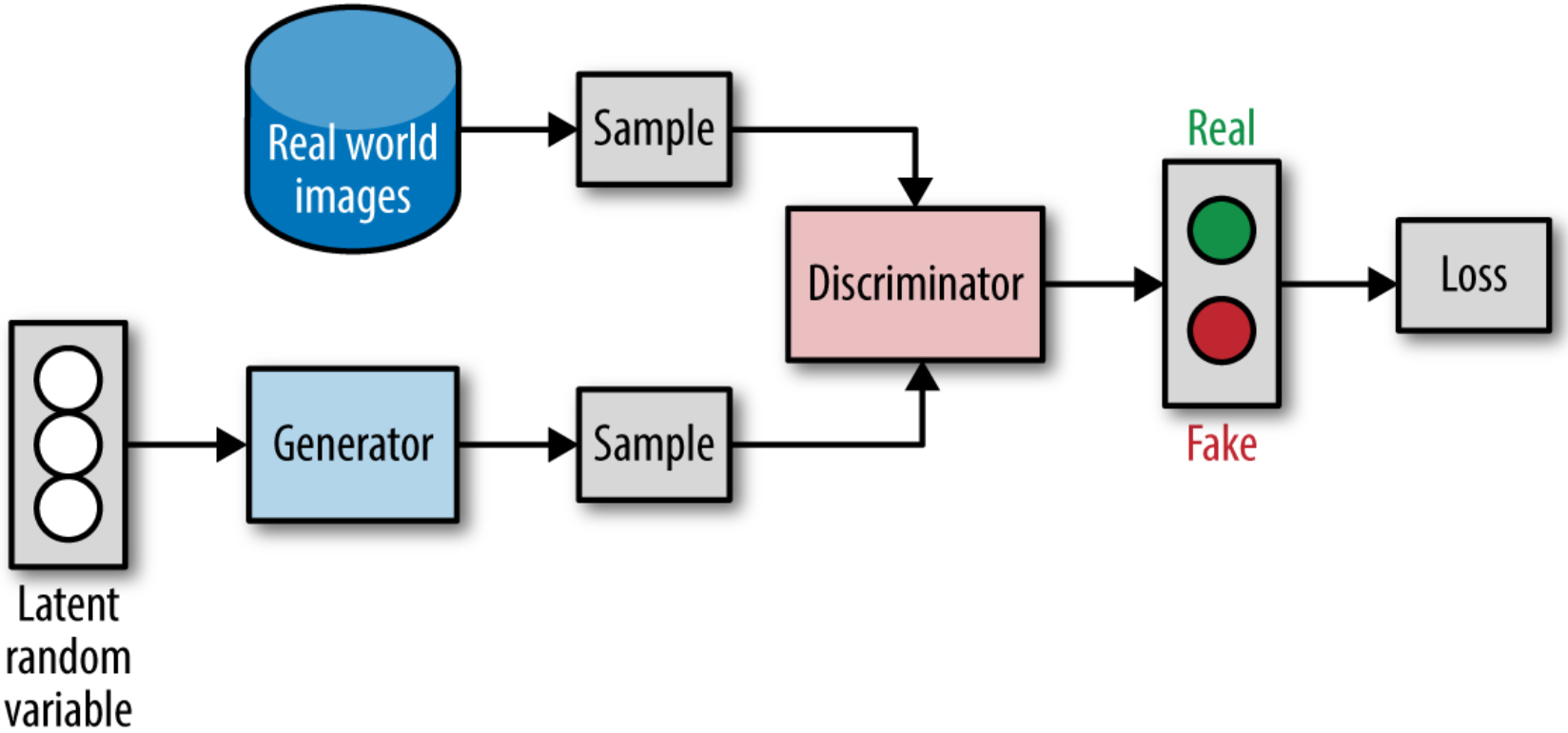}
\caption{Generative adversarial networks (GAN) architecture.
Reproduced from \cite{ramsundar:tensor:flow:deep:learning:2018} with permission of O'Reilly Media}
\label{figure:gan:architecture}
\end{figure}

An interesting example of architectural pattern
is the concept of {\em Generative Adversarial Networks} (GAN) \cite{goodfellow:gan:arxiv:2014},
as illustrated in Figure~\ref{figure:gan:architecture}.
The idea is to simultaneously train two neural networks\index{Neural!network}:

\begin{itemize}

\item a {\em generative model} (or {\em generator\index{Generator}}) G,
whose objective is to transform a random noise vector into a synthetic (faked) {\em sample\index{Sample}},
which resembles real samples drawn from a distribution of real content (images, melodies\ldots); and

\item a {\em discriminative model} (or {\em discriminator\index{Discriminator}}) D,
which estimates\index{Estimation} the probability that a sample came from the real data rather than from the generator G.

\end{itemize}

The generator is then able to produce user-appealing synthetic samples from noise vectors.

\label{section:system:midinet}

\begin{figure}
\includegraphics[width=\textwidth]{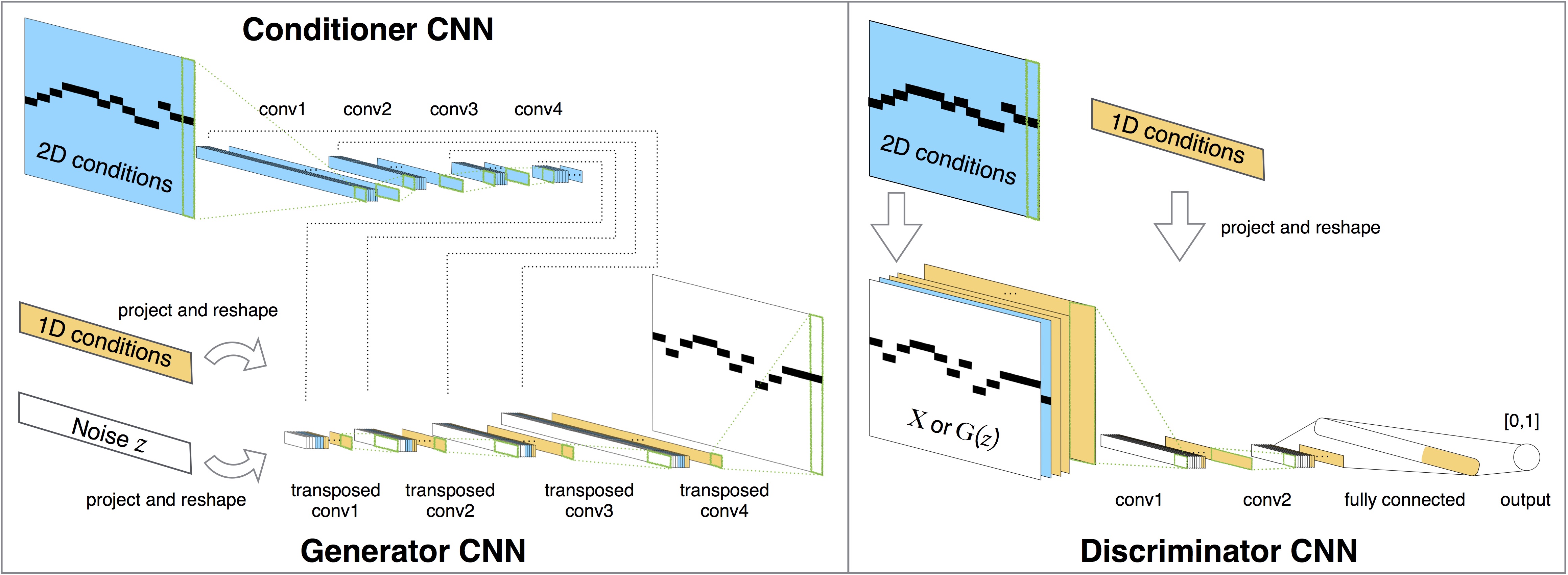}
\caption{MidiNet architecture.
Reproduced from \cite{yang:midinet:ismir:2017} with permission of the authors}
\label{figure:midinet:architecture}
\end{figure}

\begin{figure}
\includegraphics[width=\textwidth]{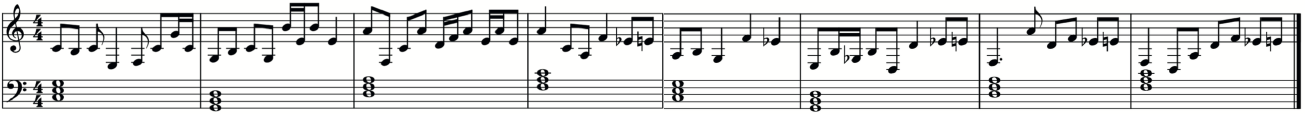}
\caption{Example of melody and chords generated by MidiNet.
Reproduced from \cite{yang:midinet:ismir:2017} with permission of the authors}
\label{figure:midinet:example}
\end{figure}

An example of the use of GAN for generating music is the MidiNet\index{MidiNet} system \cite{yang:midinet:ismir:2017},
aimed at the generation of single or multitrack pop\index{Pop} music melodies.
The architecture,
illustrated in Figure~\ref{figure:midinet:architecture},
follows
two
patterns:
{\em adversarial} (GAN)
and {\em conditional} (on history and on chords to condition melody generation)\footnote{Please refer to
	\cite{yang:midinet:ismir:2017}
	or \cite[Section~6.10.3.3]{briot:dlt4mg:springer:2019}
	for more details about this sophisticated architecture.}.
It is also {\em convolutional} (both the generator and the discriminator are convolutional networks).
The representation chosen is obtained by transforming each channel of MIDI files
into a one-hot encoding of 8 measures long piano roll representations.
Generation takes place following an iterative strategy, by sampling one measure after one measure until reaching 8 measures.
An example of generation is shown in Figure~\ref{figure:midinet:example}.

\subsection{Sampling Strategy}
\label{section:challenges:strategies:sampling}

We have discussed in the Section~\ref{section:strategy:iterative:sampling:feedforward}
the use of sampling at the output of a recurrent network in order to ensure content variability.
But sampling could also be used as the principal strategy for generation, as we will see in the two following examples.
The idea is to consider
{\em incremental variable instantiation},
where a global
representation is
incrementally instantiated by progressively refining the values of variables (e.g., pitch and duration of notes).
The main advantage is that it is possible to generate or to {\em regenerate} only an {\em arbitrary part} of the
musical content,
for a specific {\em time interval}
and/or for a specific {\em subset of tracks/voices},
without having to
regenerate the whole content.

\label{section:system:deepbach}

\begin{figure}
\includegraphics[width=0.55\textwidth]{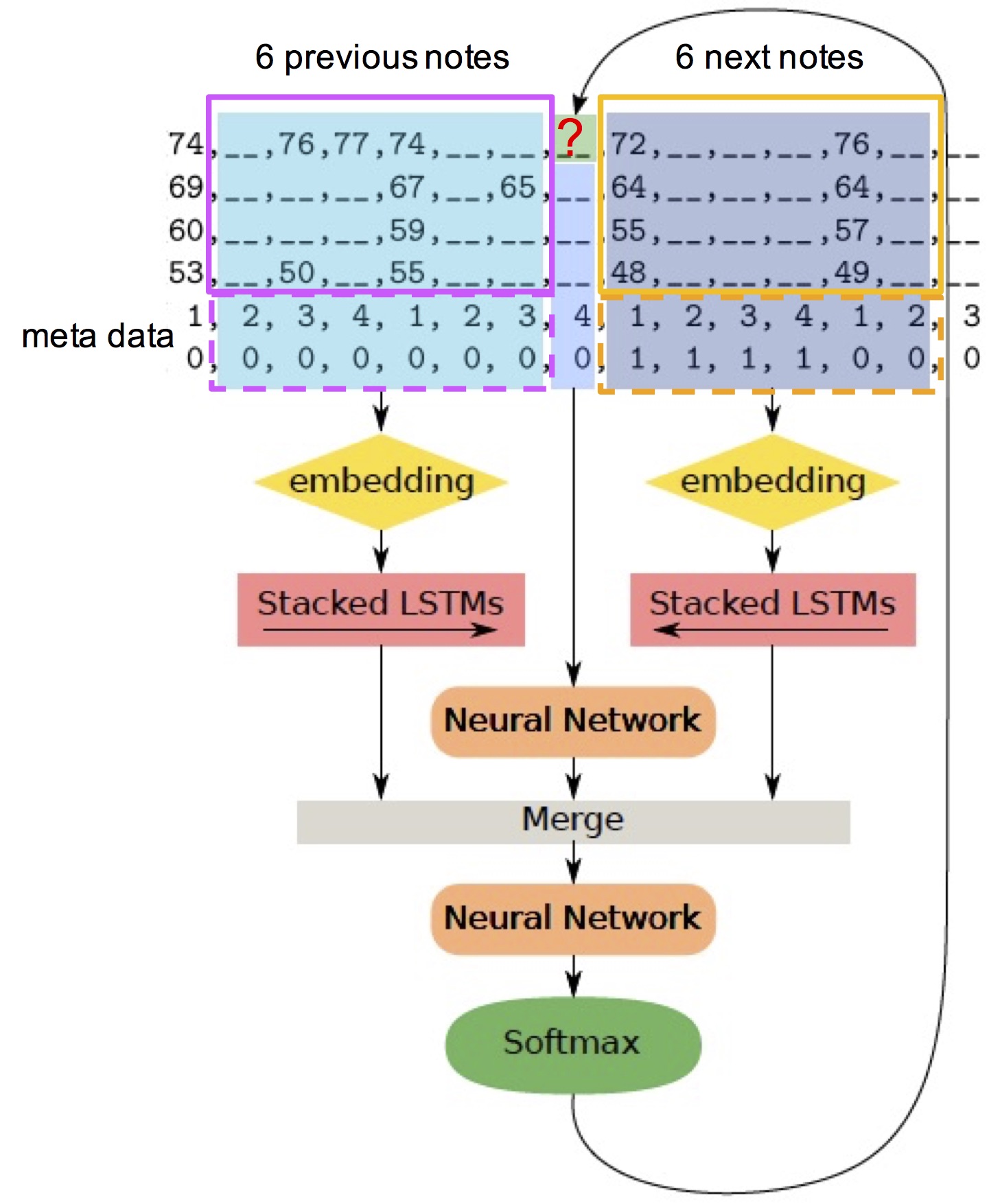}
\caption{DeepBach architecture for the soprano voice prediction.
Reproduced from \cite{hadjeres:deep:bach:arxiv:2017} with permission of the authors}
\label{figure:deep:bach:architecture:sampling}
\end{figure}

This incremental instantiation strategy has been used in the DeepBach\index{DeepBach}
architecture
\cite{hadjeres:deep:bach:arxiv:2017}
for generation of Bach\index{Bach} chorales\index{Chorale}.
The compound architecture\footnote{Actually this architecture is replicated\index{Replicate} 4 times,
	one for each voice\index{Voice} (4 in a chorale).},
shown at
Figure~\ref{figure:deep:bach:architecture:sampling},
combines two recurrent
and two feedforward networks\index{Feedforward network}.
As opposed to standard use of recurrent networks, where a single time direction is considered,
DeepBach architecture considers the two directions {\em forward\index{Forward}} in time and {\em backward\index{Backwards}} in time.
Therefore, two recurrent networks (more precisely, LSTM)
are used,
one summing up past information and another summing up information coming from the future,
together with a non recurrent network for notes occurring at the same time.
Their three outputs are merged and passed as the input of a final feedforward neural network,
whose output is the estimated distribution for all notes time slices for a given voice.
The first 4 lines\footnote{The two bottom lines
	correspond to metadata\index{Metadata} (fermata\index{Fermata} and beat\index{Beat} information), not detailed here.}
of the example data on top of the Figure~\ref{figure:deep:bach:architecture:sampling}
correspond to the 4 voices\index{Voice}.

Training, as well as generation, is not done in the conventional way
for neural networks.
The objective is to predict the value of current note for a a given voice
(shown
with a red
``?''
on top center of
Figure~\ref{figure:deep:bach:architecture:sampling}),
using as information surrounding contextual notes.
The training set is formed on-line by repeatedly randomly selecting a note in a voice from an example of the corpus
and its surrounding context.
Generation\index{Generation} is done
by sampling\index{Sampling},
using a pseudo-Gibbs sampling\index{Pseudo-Gibbs sampling} incremental and iterative algorithm
(see details in \cite{hadjeres:deep:bach:arxiv:2017})
to
update values (each note) of a polyphony, following the distribution that the network has learnt. 

\label{section:interactivity:deepbach}

\begin{figure}
\includegraphics[width=0.7\textwidth]{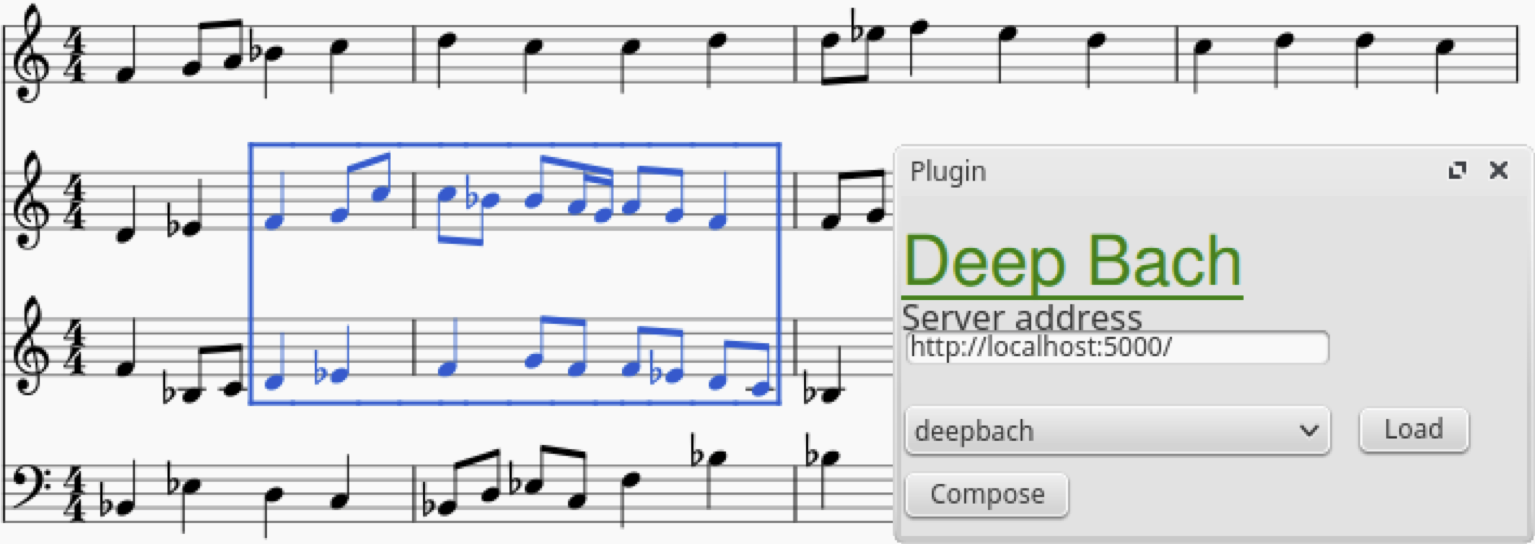}
\caption{DeepBach user interface. Selecting an interval and two voices
(alto and tenor)
to be regenerated.
Reproduced from \cite{hadjeres:deep:bach:arxiv:2017} with permission of the authors}
\label{figure:deep:bach:user:interface}
\end{figure}

The advantage of this method is that generation may be tailored.
For example, if the user makes some local adjustment,
he can resample only some of the corresponding counterpoint voices (e.g., alto and tenor)
for the chosen interval (e.g., a measure and a half),
as shown in Figure~\ref{figure:deep:bach:user:interface}.


\label{section:system:cononet}

\begin{figure}
\includegraphics[width=0.8\textwidth]{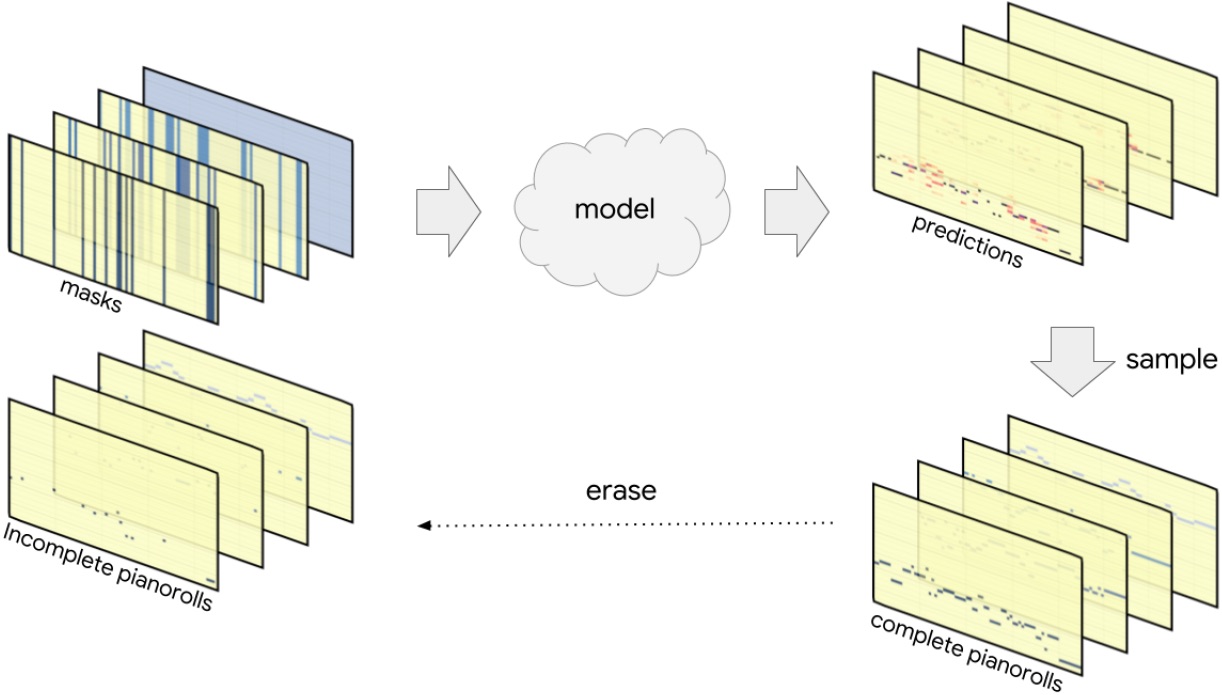}
\caption{Coconet Architecture. Reproduced from \cite{huang:coconet:model:bach:doodle:magenta:2019} with permission of the authors}
\label{figure:coconet:architecture}
\end{figure}

Coconet \cite{huang:counterpoint:convolution:ismir:2017},
the architecture used for implementing the Bach Doodle
(introduced in Section~\ref{section:system:bach:doodle}),
is another example of this approach.
It uses a Block Gibbs sampling algorithm for generation
and a different architecture (using masks to indicate for each time slice whether the pitch for that voice is known,
see Figure~\ref{figure:coconet:architecture}).
Please refer to \cite{huang:counterpoint:convolution:ismir:2017}
and \cite{huang:coconet:model:bach:doodle:magenta:2019} for details.
An example of counterpoint accompaniment generation has been shown in Figure~\ref{figure:bach:doodle}.

\subsection{Creation by Refinement Strategy}
\label{section:challenges:strategies:control:input:manipulation}
\label{section:strategy:input:manipulation}

Lewis' creation by refinement strategy has been introduced in Section~\ref{section:system:lewis}.
It has been ``reinvented'' by various systems such as Deep Dream and DeepHear,
as discussed in Section~\ref{section:system:lewis:influence}.

\label{section:systems:c-rbm}

An example of application to music
is the generation algorithm for the C-RBM architecture
\cite{lattner:structure:polyphonic:generation:jcms:2018}.
The architecture is a refined (convolutional\footnote{The architecture is convolutional (only)
	on the time\index{Time} dimension,
	in order to model temporally invariant\index{Invariant} motives\index{Motif},
	but not pitch\index{Pitch} invariant motives
	which would break the notion of tonality\index{Tonality}.})
	restricted Boltzmann machine\index{Restricted Boltzmann machine}
(RBM\index{RBM}\footnote{Because of space limitation,
	and the fact that RBMs are not mainstream,
	we do not detail here the characteristics of RBM
	(see, e.g., \cite[Section~20.2]{goodfellow:deep:learning:book:2016} or \cite[Section~5.7]{briot:dlt4mg:springer:2019} for details).
	In a first approximation for this paper,
	we may consider an RBM as analog to an autoencoder,
	except
	two differences:
	the input and output layers are merged (and named the visible layer)
	and the model is stochastic.}).
It is trained to learn the
{\em local structure} (musical texture/style)
of a corpus\index{Corpus} of music (in this case, Mozart\index{Mozart} sonatas\index{Sonata}).
The main idea is to impose
onto (and {\em during}) the creation of a new musical content
some {\em global\index{Global} structure\index{Structure}}
seen as a {\em structural template\index{Template}} from an existing reference musical piece\footnote{This is named
	{\em structure imposition},
	with the same basic approach that of style transfer \cite{dai:music:style:transfer:arxiv:2018},
	except that of a high-level structure.}.
The global structure is expressed through three types of
constraints\index{Constraint}:

\begin{itemize}

\item {\em self-similarity}, to specify a {\em global structure\index{Structure}} (e.g., AABA) in the generated music piece.
This is modeled by minimizing the distance\index{Distance}
between the self-similarity\index{Similarity}\index{Self-similarity}
matrices of the reference target and of the intermediate solution;

\item {\em tonality constraint}, to specify a {\em key\index{Key}} (tonality\index{Tonality}).
To control the key in a given temporal\index{Temporal} window\index{Window},
the distribution\index{Distribution} of pitch classes\index{Pitch!class}
is compared with the
key profiles\index{Profile}
of the reference; and

\item {\em meter constraint}, to impose a specific {\em meter\index{Meter}}
(also named a {\em time signature\index{Time!signature}},
e.g., 4/4)
and its related rhythmic\index{Rhytmic} pattern\index{Pattern}
(e.g.,
an accent\index{Accent} on the
third beat\index{Beat}).
The relative occurrence of note onsets
within a measure is constrained to follow that of the reference.

\end{itemize}

\begin{figure}
\includegraphics[width=\textwidth]{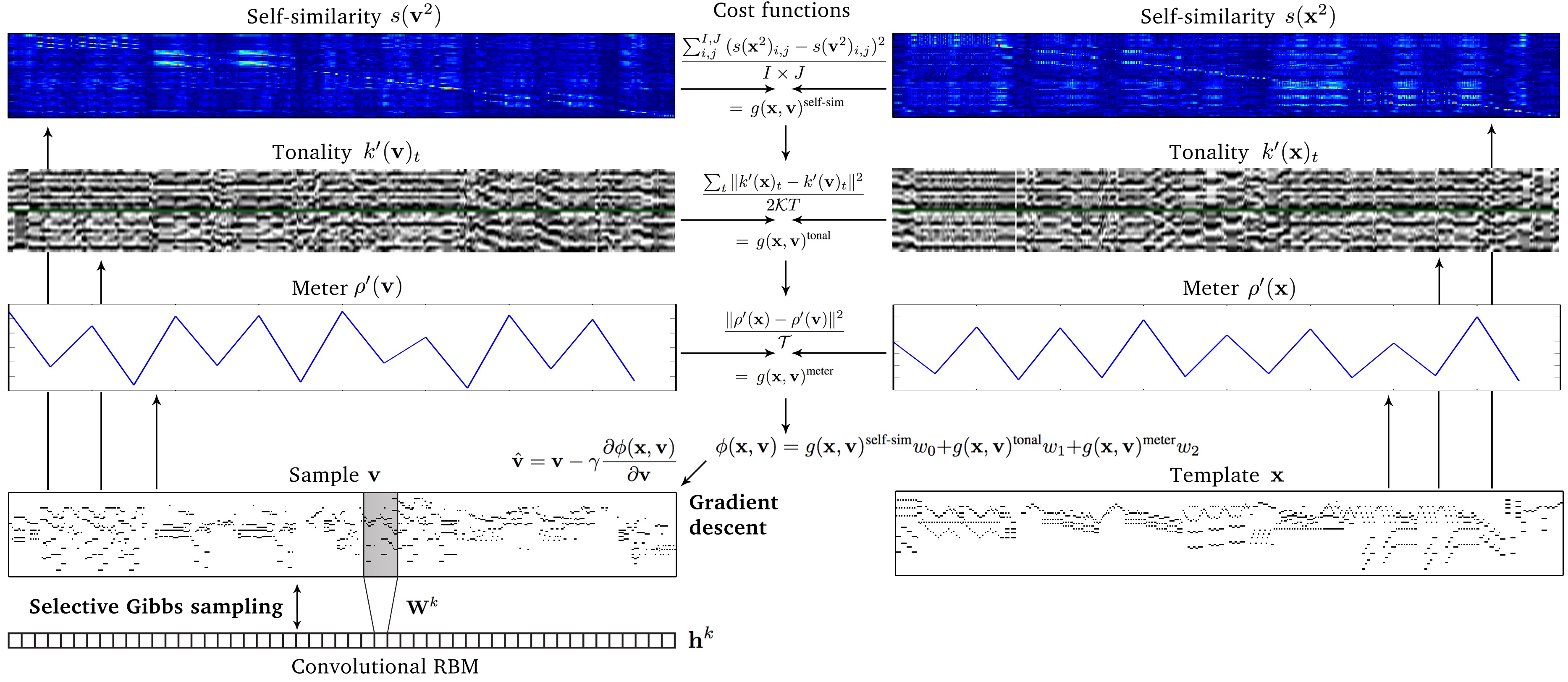}
\caption{C-RBM Architecture generation algorithm. Reproduced from \cite{lattner:structure:polyphonic:generation:jcms:2018}
with permission of the authors}
\label{figure:crbmc:architecture}
\end{figure}

\begin{figure}
\includegraphics[width=0.8\textwidth]{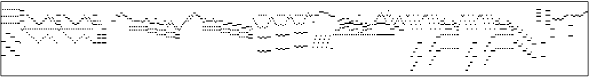}
\includegraphics[width=0.8\textwidth]{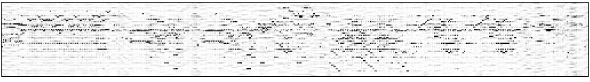}
\includegraphics[width=0.8\textwidth]{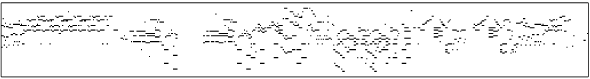}
\caption{Illustration of constrained sampling.
Piano roll representation of:
(top) Template piece, (middle) Intermediate sample after the GD phase, (bottom) Sample after the GS phase.
Reproduced from \cite{lattner:structure:polyphonic:generation:jcms:2018} with permission of the authors}
\label{figure:constrained:sampling:example:simplified}
\end{figure}

Generation\index{Generation} is performed via {\em constrained sampling\index{Constrained sampling}},
a mechanism to restrict the set of possible solutions in the sampling\index{Sampling} process
according to some pre-defined constraints\index{Constraint}.
The principle of the process (illustrated at Figure~\ref{figure:crbmc:architecture}) is as follows.
At first, a sample is randomly\index{Randomly} initialized\index{Initialize}, following the standard uniform distribution\index{Distribution}.
A step of constrained sampling\index{Constrained sampling}
is composed of
$n$ runs of gradient descent\index{Gradient descent} (GD)
to impose the high-level structure,
followed by $p$ runs of {\em selective Gibbs sampling\index{Selective Gibbs sampling}} (GS)
to selectively realign the sample onto the learnt distribution.
A simulated annealing\index{Simulated annealing} algorithm is applied in order to control exploration to favor good solutions.
Figure~\ref{figure:constrained:sampling:example:simplified} shows an example of a generated sample in piano roll format.

\subsection{Other Architectures and Strategies}
\label{section:architecture:other}

Researchers in the domain of deep learning techniques for music generation
are designing and experimenting with various architectures and strategies\footnote{This paper is obviously not exhaustive.
	Interested readers
	may refer, e.g., to \cite{briot:dlt4mg:springer:2019} for
	additional examples and details.},
in most cases combinations or refinements of existing ones,
or sometimes with novel types, as, e.g., in the case of MusicTransformer \cite{huang:music:transformer:arxiv:2018}.
However,
there is no guarantee that combining a maximal variety of types will make a sound and accurate architecture\footnote{As in the case of a good cook,
	whose aim is not to simply mix {\em all} possible ingredients
	but to discover original successful combinations.}.
Therefore, it is important to continue to deepen our understanding and to explore solutions
as well as their possible articulations and combinations.
We hope that this paper could contribute to that objective.

\section{Open Issues and Trends}
\label{section:open:issues}

Because of space limitation, we will only sketch some of open issues and current trends about using neural networks and deep learning techniques
for music generation.
Following (and further developed in) \cite{mgbdlcd:ncaa:2018},
we consider some main challenges to be:
control, structure,
creativity and interactivity.

\begin{itemize}

\item {\em Control} is necessary to inject constraints (e.g., tonality, rhythm) in the generation,
as witnessed by the C-RBM architecture (see Section~\ref{section:systems:c-rbm}).
Some challenge is that a deep learning architecture is a kind of black box,
therefore some control entry points (hooks) need to be identified,
such as:
the {\em input} (in the case of creation by refinement,
as introduced in Section~\ref{section:system:lewis},
or by using an extra conditioning input,
as in Anticipation-RNN \cite{hadjeres:anticipation:rnn:arxiv:2017});
the {\em output} (in the case of constrained sampling, as in C-RBM\footnote{Actually,
	in that case of an RBM,
	the output is also the input.}); or
an {\em encapsulation} (in the case of a reformulation through reinforcement learning,
as in RL Tuner \cite{jaques:rl:tuner:arxiv:2016}).

\item {\em Structure} is an important issue. Music generated by a deep learning architecture may be very pleasing for less than a minute
but usually starts to be boring after a little while because of the absence of a clear sense of direction.
Structure imposition is a first direction, as in C-RBM, or by using hierarchical architectures as in Music-VAE.
A more ambitious direction is to favor the emergence of structure.

\item {\em Creativity} is obviously a desired objective, while a difficult and profound issue.
Neural networks are actually not well prepared, 
as their strength in generating music very conformant to a style learnt turns out to be a weakness for originality.
Therefore, various strategies are being explored to try to favor exiting from the ``comfort zone'',
without losing it all.
A notable attempt has been proposed for creating paintings in \cite{elgammal:can:arxiv:2017},
by extending a GAN architecture to favor the generation of content difficult to classify within
existing styles, and therefore favoring the emergence of new styles.
Meanwhile, as discussed in Section~\ref{section:introduction:motivation:assistance:versus},
we believe that it is more interesting to use deep learning architectures to assist human musicians to create and construct music,
than pursuing purely autonomous music generating systems.

\item Therefore,  {\em interactivity} is necessary to be able to allow a musician to incrementally develop a creation
with the help of a deep learning-based system.
This could be done, e.g., by allowing partial regeneration\footnote{Like inpainting for the regeneration
	of missing or deteriotating parts of images.},
as in \cite{pati:musical:score:inpainting:arxiv:2019},
or by focusing on an incremental sampling strategy, as described in Section~\ref{section:challenges:strategies:sampling}.

\end{itemize}

\section{Conclusion}
\label{section:conclusion}

The use of artificial neural networks and deep learning architectures and techniques for the generation of music (as well as other artistic content)
is a very active area of research.
In this paper, we have:
introduced the domain;
analyzed early and pioneering proposals;
and introduced a conceptual framework to help at analyzing and classifying the large diversity of
architectures
and experiments
described in the literature,
while illustrating it by various
examples.
We hope that this paper will help in better
understanding the domain and trends of deep learning-based music generation.


\subsection*{Acknowledgements}

This paper is dedicated to the memory of my beloved mother.

We thank Ga\"etan Hadjeres and Fran\c{c}ois Pachet for their participation to the book \cite{briot:dlt4mg:springer:2019}
which has been a significant initial input for this paper;
CNRS,
Sorbonne Universit\'e,
UNIRIO
and PUC-Rio
for their support;
and the various participants of our course on the topic\footnote{Course material
	is available online at: {\tt http://www-desir.lip6.fr/{\textasciitilde}briot/cours/unirio3/}.}
for their feedback.

\bibliographystyle{spmpsci}      
\bibliography{nn4music}   

\section*{Annex -- Glossary}

\begin{description}

\item[\bf Activation function]
The function applied to the weighted sum for each neuron of a given layer.
It is usually nonlinear
(to introduce nonlinearity,
in order
to address the linear separability limitation of the Perceptron).
Common examples are sigmoid or ReLU.
The activation function of the output layer is a specific case (see Output layer activation function).

\item[\bf Algorithmic composition\index{Algorithmic composition}]
The use of algorithms and computers
to generate music compositions (symbolic form) or music pieces (audio form).
Examples of models and algorithms are: grammars, rules,
stochastic processes (e.g., Markov chains), evolutionary methods
and artificial neural networks.

\item[\bf Architecture\index{Architecture}]
An (artificial neural network) architecture is the structure of the organization of computational units (neurons), usually grouped in layers, and their weighted connexions.
Examples of types of architecture are: feedforward (aka multilayer Perceptron), recurrent (RNN), autoencoder and generative adversarial networks (GAN).
Architectures process encoded representations (in our case of a musical content).

\item[\bf Artificial neural network\index{Artificial!neural network}]
A family of bio-inspired machine learning algorithms whose model is based on weighted connexions between computing units (neurons).
Weights are incrementally adjusted during the training phase in order for the model to fit the data (training examples).

\item[\bf Attention mechanism\index{Attention mechanism}]
A mechanism inspired by the human visual system which focuses at each time step on some specific elements of the input sequence.
This is modeled by weighted connexions onto the sequence elements (or onto the sequence of hidden units)
which are subject to be learned.

\item[\bf Autoencoder\index{Autoencoder}]
A specific case of artificial neural network architecture with an output layer mirroring the input layer and with one hidden layer.
Autoencoders are good at extracting features.

\item[\bf Backpropagation\index{Backpropagation}]
A short hand for ``backpropagation of errors'',
it is the algorithm used to compute the gradients
of the cost function.
Gradients will be used to guide the minimization of the cost function in order to fit the data.


\item[\bf Bias\index{Bias}]
The $b$ offset term of a simple linear regression model $h(\text{x}) = b + \theta \text{x}$
and by extension of a neural network layer.

\item[\bf Bias node\index{Bias!node}]
The node of a neural network layer corresponding to a bias.
Its constant value is 1 and is usually notated as $+1$.



\item[\bf Challenge\index{Challenge}]
One of the qualities requirements that may be desired for music generation.
Examples of challenges are: incrementality, originality and structure.


\item[\bf Classification\index{Classification}]
A machine learning task about the attribution of an instance to a class (from a set of possible classes).
An example is to determine if next note is a C$_4$, a C$\sharp_4$, etc.

\item[\bf Compound architecture\index{Compound architecture}]
An artificial neural network architecture which is the result of some combination of some architectures.
Examples of types of combination are composition, nesting and pattern instantiation.

\item[\bf Conditioning architecture\index{Conditioning!architecture}]
The parametrization of an artificial neural network architecture by some conditioning information
(e.g., a bass line, a chord progression\ldots)
represented via a specific extra input,
in order to guide the generation.

\item[\bf Connexion\index{Connexion}]
A relation between a neuron and another neuron representing a computational flow from the output of the first neuron
to an input of the second neuron.
A connexion is modulated by a weight which will be adjusted during the training phase.


\item[\bf Convolution\index{Convolution}]
In mathematics, a mathematical operation on two functions sharing the same domain
that produces a third function which is the integral (or the sum in the discrete case -- the case of images made of pixels)
of the pointwise multiplication of the two functions varying within the domain in an opposing way.
Inspired both by mathematical convolution and by a model of human visions,
it has been adapted to artificial neural networks and it improves pattern recognition accuracy
by exploiting the spatial local correlation present in natural images.
The basic principle is to slide a matrix (named a filter, a kernel or a feature detector)
through the entire image (seen as the input matrix),
and for each mapping position to compute the dot product of the filter with each mapped portion of the image and
then sum up all elements of the resulting matrix.

\item[\bf Correlation\index{Correlation}]
Any statistical relationship, whether causal or not, between two random variables.
Artificial neural networks are good at extracting correlations between variables, for instance between input variables and output variables
and also between input variables.

\item[\bf Cost function (aka Loss function)]
The function used for measuring the difference between the prediction by an artificial neural network architecture (\^y) and the actual target (true value y).
Various cost functions may be used, depending on the task (prediction or classification) and the encoding of the output, e.g.,
mean squared error, binary cross-entropy and categorical cross entropy.

\item[\bf Counterpoint\index{Counterpoint}]
In musical theory, an approach for the accompaniment of a melody through a set of other melodies (voices).
An example is a chorale with 3 voices (alto, tenor and bass) matching a soprano melody.
Counterpoint focuses on the horizontal relations between successive notes
for each simultaneous melody (voice)
and then considers the vertical relations between their progression (e.g., to avoid parallel fifths).

\item[\bf Creation by refinement strategy\index{Input!manipulation strategy}]
A strategy for generating content based on the incremental modification of a representation to be processed
by an artificial neural network architecture.

\item[\bf Cross-entropy\index{Cross-entropy}]
A function measuring the dissimilarity between two probability distributions.
It is used as a cost (loss) function for a classification task
to measure the difference between the prediction by an artificial neural network architecture (\^y) and the actual target (true value y).
There are two types of cross-entropy cost functions:
binary cross-entropy when the classification is binary
and categorical cross-entropy when the classification is multiclass with a single label
to be selected.


\item[\bf Dataset\index{Dataset}]
The set of examples used for training an artificial neural network architecture.

\item[\bf Decoder\index{Decoder}]
The decoding component of an autoencoder which reconstructs the compressed representation (an embedding) from the hidden layer
into a representation at the output layer as close as possible to the initial data representation at the input layer.

\item[\bf Decoder feedforward strategy\index{Decoder!feedforward strategy}]
A strategy for generating content based on an autoencoder architecture
in which values are assigned onto the latent variables of the hidden layer and forwarded into the decoder component of the architecture in order to generate
a musical content corresponding to the abstract description inserted.

\item[\bf Deep learning\index{Deep!learning} (aka Deep neural networks\index{Deep!neural network})]
An artificial neural network architecture with a significant number of successive layers.


\item[\bf Discriminator\index{Discriminator}]
The discriminative model component of generative adversarial networks (GAN)
which estimates the probability that a sample came from the real data rather than from the generator.

\item[\bf Disentanglement]
The objective of separating the different factors governing variability in the data
(e.g., in the case of human images,
identity of the individual and facial expression,
in the case of music,
note pitch range and note duration range).


\item[\bf Embedding\index{Embedding}]
In mathematics, an injective and structure-preserving mapping.
Initially used for natural language processing,
it is now often used in deep learning as a general term for encoding
a given representation into a vector representation.

\item[\bf Encoder\index{Encoder}]
The encoding component of an autoencoder which transforms the data representation from the input layer
into a compressed representation (an embedding) at the hidden layer.

\item[\bf Encoding\index{Encoding}]
The encoding of a representation
consists in the mapping of the representation
(composed of a set of variables, e.g., pitch or dynamics)
into a set of inputs (also named input nodes or input variables)
for the neural network architecture.
Examples of encoding strategies are: value encoding, one-hot encoding and many-hot encoding.

\item[\bf End-to-end architecture\index{End-to-end architecture}]
An artificial neural network architecture that processes the raw unprocessed data -- without any pre-processing, transformation of representation,
or extraction of features -- to produce a final output.







\item[\bf Feedforward\index{Feedforward}]
The basic way for a neural network architecture to process an input by feedforwarding
the input data into the successive layers of neurons of the architecture until producing the output.
A feedforward neural architecture (also named multilayer neural network or multilayer Perceptron, MLP)
is composed of successive layers, with at least one hidden layer.

\item[\bf Fourier transform\index{Fourier transform}]
A transformation (which could be continuous or discrete)
of a signal into the decomposition into its elementary components (sinusoidal waveforms).
As well as compressing the information, its role is fundamental for musical purposes
as it reveals the harmonic components of the signal.

\item[\bf Generative adversarial networks\index{Generative!adversarial networks} (GAN\index{GAN})]
A compound architecture composed of two component architectures, the generator and the discriminator,
who are trained simultaneously with opposed objectives.
The objective of the generator is to generate synthetic samples resembling real data
while the objective of the discriminator is to detect synthetic samples.

\item[\bf Generator\index{Generator}]
The generative model component of generative adversarial networks (GAN)
whose objective is to transform a random noise vector into a synthetic (faked) sample
which resembles real samples drawn from a distribution of real data.

\item[\bf Gradient\index{Gradient}]
A partial derivative of the cost function with respect to a weight parameter or a bias.

\item[\bf Gradient descent\index{Gradient!descent}]
A basic algorithm for training a linear regression model and an artificial neural network.
It consists in an incremental update of the weight (and bias) parameters guided by the gradients
of the cost function
until reaching a minimum.

\item[\bf Harmony\index{Harmony}]
In musical theory, a system for organizing simultaneous notes.
Harmony focuses on the vertical relations between simultaneous notes,
as objects on their own (chords),
and then considers the horizontal relations between them (e.g., harmonic cadences).

\item[\bf Hidden layer\index{Hidden!layer}]
Any neuron layer located between the input layer and the output layer of a neural network architecture.

\item[\bf Hold\index{Hold}]
The information about a note that extends its duration over a single time step. 

\item[\bf Hyperparameter\index{Hyperparameter}]
Higher-order parameters about the configuration of a neural network architecture and its behavior.
Examples are: number of layers, number of neurons for each layer, learning rate and stride (for a convolutional architecture).

\item[\bf Input layer\index{Input!layer}]
The first layer of a neural network architecture.
It is an interface consisting in a set of nodes without internal computation.


\item[\bf Iterative feedforward strategy\index{Iterative feedforward strategy}]
A strategy for generating content by generating its successive time slices.

\item[\bf Latent variable\index{Latent!variable}]
In statistics, a variable which is not directly observed.
In deep learning architectures, variables within a hidden layer.
By sampling a latent variable(s), one may control the generation, e.g., in the case of a variational autoencoder.

\item[\bf Layer\index{Layer}]
A component of a neural network architecture composed of a set of neurons.

\item[\bf Linear regression\index{Linear!regression}]
Regression for an assumed linear relationship
between a scalar variable and one or several explanatory variable(s).

\item[\bf Linear separability\index{Linear!separability}]
The ability to separate by a line or a hyperplane the elements of two different classes represented in an Euclidian space.

\item[\bf Long short-term memory\index{Long!short-term memory} (LSTM\index{LSTM})]
A type of recurrent neural network (RNN) architecture with capacity for learning long term correlations
and not suffering from the vanishing or exploding gradient problem
during the training phase
(in backpropagation through time,
recurrence brings repetitive multiplications
which could lead to amplify numerical errors).
The idea is to secure information in memory cells protected from the standard data flow of the recurrent network.
Decisions about writing to, reading from and forgetting the values of cells
are performed by the opening or closing of gates
and are expressed at a distinct control level, while being learnt during the training process.

\item[\bf Many-hot encoding\index{Many-hot encoding}]
Strategy used to encode simultaneously several values of a categorical variable,
e.g., a triadic chord composed of three note pitches.
As for a one-hot encoding, it is based on a vector having as its length the number of possible values (e.g., from C$_4$ to B$_4$).
Each occurrence of a note is represented with a corresponding 1 with all other elements being 0.

\item[\bf Markov chain\index{Markov!chain}]
A stochastic model describing a sequence of possible states.
The chance to change from the current state to a state or to another state is governed by a probability
and does not depend on previous states.

\item[\bf Melody\index{Melody}]
The abbreviation of a single-voice monophonic melody,
that is a sequence of notes for a single instrument
with at most one note at the same time.

\item[\bf Musical instrument digital interface\index{Musical!instrument digital interface} (MIDI\index{MIDI})]
A technical standard that describes a protocol, a digital interface and connectors for interoperability
between various electronic musical instruments, softwares and devices.

\item[\bf Multilayer Perceptron\index{Multilayer!Perceptron} (MLP\index{MLP})]
A feedforward neural architecture composed of successive layers,
with at least one hidden layer. Also named Feedforward architecture.

\item[\bf Multivoice\index{Multivoice} (aka Multitrack\index{Multitrack})]
The abbreviation of a multivoice polyphony,
that is a set of sequences of notes intended for more than one voice or instrument.

\item[\bf Neuron\index{Neuron}]
The atomic processing element (unit) of an artificial neural network architecture.
A neuron
has several input connexions, each one with an associated weight,
and one output.
A neuron will compute the weighted sum of all its input values and then apply the activation function associated to its layer
in order to compute its output value.
Weights will be adjusted during the training phase of the neural network architecture.

\item[\bf Node\index{Node}]
The atomic structural element of an artificial neural network architecture.
A node could be a processing unit (a neuron) or a simple interface element for a value, e.g., in the case of the input layer or a bias node.


\item[\bf Objective\index{Objective}]
The nature and the destination of the musical content to be generated by a neural network architecture.
Examples of objectives are: a monophonic melody to be played by a human flutist
and a polyphonic accompaniment played by a synthesizer.

\item[\bf One-hot encoding\index{One-hot encoding}]
Strategy used to encode a categorical variable (e.g., a note pitch)
as a vector having as its length the number of possible values (e.g., from C$_4$ to B$_4$).
A given element (e.g., a note pitch) is represented with a corresponding 1 with all other elements being 0.
The name comes from digital circuits, one-hot referring to a group of bits among which the only legal (possible) combinations
of values are those with a single high (hot) (1) bit, all the others being low (0).

\item[\bf Output layer\index{Output!layer}]
The last layer of a neural network architecture.

\item[\bf Output layer activation function]
The activation function of the output layer,
which is usually: identity for a prediction task,
sigmoid for a binary classification task
and softmax for a multiclass single-label classification task.



\item[\bf Parameter\index{Parameter}]
The parameters of an artificial neural network architecture
are the weights associated to each connexion between neurons
as well as the biases associated to each layer.

\item[\bf Perceptron\index{Perceptron}]
One of the first artificial neural network architecture, created by Rosenblatt in 1957.
It had no hidden layer and suffered from the linear separability limitation.

\item[\bf Piano roll\index{Piano!roll}]
Representation of a melody (monophonic or polyphonic) inspired from automated pianos.
Each ``perforation'' represents a note control information, to trigger a given note.
The length of the perforation corresponds to the duration of a note.
In the other dimension, the localization (height) of a perforation corresponds to its pitch.

\item[\bf Pitch class\index{Pitch!class}]
 The name of the corresponding note (e.g., C) independently of the octave position.
Also named chroma.

\item[\bf Polyphony\index{Polyphony}]
The abbreviation of a single-voice polyphony, that is a sequence of notes for a single instrument
(e.g., a guitar or a piano)
with possibly simultaneous notes.



\item[\bf Prediction]
See regression.


\item[\bf Recurrent connexion\index{Recurrent!connexion}]
A connexion from an output of a node to its input.
By extension, the recurrent connexion of a layer fully connects the outputs of all its nodes to all inputs of all its nodes.
This is the basis of a recurrent neural network (RNN) architecture.

\item[\bf Recurrent neural network\index{Recurrent!neural network} (RNN\index{RNN})]
A type of artificial neural network architecture with recurrent connexions and memory.
It is used to learn sequences.

\item[\bf Recursive feedforward strategy\index{Iterative feedforward strategy}]
A special case of iterative feedforward strategy where the current output is used as the next input.

\item[\bf Regression\index{Regression}]
In statistics, regression is an approach for modeling the relationship
between a scalar variable and one or several explanatory variable(s).

\item[\bf Reinforcement learning\index{Reinforcement!learning}]
An area of machine learning concerned with an agent making successive decisions about an action in an environment
while receiving a reward (reinforcement signal) after each action.
The objective for the agent is to find the best policy maximizing its cumulated rewards.

\item[\bf Reinforcement strategy\index{Reinforcement!strategy}]
A strategy for content generation by modeling generation of successive notes as a reinforcement learning problem
while using an RNN as a reference for the modeling of the reward.
Therefore, one may introduce arbitrary control objectives (e.g., adherence to current tonality, maximum number of repetitions, etc.)
as additional reward terms.

\item[\bf ReLU\index{ReLU}]
The rectified linear unit function, which may be used as a hidden layer nonlinear activation function,
specially in the case of convolutions.

\item[\bf Representation\index{Representation}]
The nature and format of the information (data) used to train an architecture and to generate musical content.
Examples of types of representation are:
waveform signal, spectrum, piano roll and MIDI.

\item[\bf Requirement]
One of the qualities that may be desired for music generation.
Examples are: content variability, incrementality, originality and structure.

\item[\bf Rest\index{Rest}]
The information about the absence of a note (silence) during one (or more) time step(s). 

\item[\bf Restricted Boltzmann machine\index{Restricted Boltzmann machine} (RBM\index{RBM})]
A specific type of artificial neural network that can learn a probability distribution over its set of inputs.
It is stochastic, has no distinction between input and output, and uses a specific learning algorithm.

\item[\bf Sampling\index{Sampling}]
The action of producing an item (a sample) according to a given probability distribution over the possible values.
As more and more samples are generated, their distribution should more closely approximate the given distribution.

\item[\bf Sampling strategy\index{Sampling!strategy}]
A strategy for generating content where variables of a content representation are incrementally instantiated and refined according to a target probability distribution
which has been previously learnt.

\item[\bf Seed-based generation\index{Seed!-based generation}]
An approach to generate arbitrary content (e.g., a long melody) with a minimal (seed) information (e.g., a first note).


\item[\bf Sigmoid\index{Sigmoid}]
Also named the logistic function, it is used as an output layer activation function for binary classification tasks
and it may also be used as a hidden layer
activation function.

\item[\bf Single-step feedforward strategy\index{Single!-step feedforward strategy}]
A strategy for generating content where a feedforward architecture processes in a single processing step
a global temporal scope representation which includes all time slices.

\item[\bf Softmax\index{Softmax}]
Generalization of the sigmoid (logistic) function to the case of multiple classes.
Used as an output layer activation function for multiclass single-label classification.



\item[\bf Spectrum\index{Spectrum}]
The representation of a sound in terms of the amount of vibration at each individual (as a function of) frequency.
It is computed by a Fourier transformation	which decomposes the original signal into its elementary (harmonic) components (sinusoidal waveforms).

\item[\bf Stacked autoencoder\index{Stacked autoencoder}]
A set of hierarchically nested autoencoders
with decreasing numbers of hidden layer units.

\item[\bf Strategy\index{Strategy}]
The way the architecture will process representations in order to generate
the objective while matching desired requirements.
Examples of types of strategy are: single-step feedforward, iterative feedforward and decoder feedforward.


\item[\bf Style transfer\index{Style!transfer}]
The technique for capturing a style (e.g., of a given painting, by capturing the correlations between neurons for each layer)
and applying it onto another content.







\item[\bf Time slice\index{Time!slice}]
The time interval considered as an atomic portion (grain) of the temporal representation used by an artificial neural network architecture.

\item[\bf Time step\index{Time!step}]
The atomic increment of time considered by an artificial neural network architecture.



\item[\bf Turing test\index{Turing!test}]
Initially codified in 1950 by Alan Turing
and named by him the ``imitation game'',
the ``Turing test'' is a test of the ability for a machine to exhibit intelligent behavior equivalent to
(and more precisely, indistinguishable from) the behavior of a human.
In his imaginary experimental setting, Turing proposed the test to be a natural language conversation
between a human (the evaluator) and a hidden actor (another human or a machine).
If the evaluator cannot reliably tell the machine from the human, the machine is said to have passed the test.

\item[\bf Unit\index{Unit}]
See neuron.



\item[\bf Variational autoencoder\index{Variational!autoencoder} (VAE\index{VAE})]
An autoencoder with the added constraint that the encoded representation
(its latent variables)
follows some prior probability distribution, usually a Gaussian distribution.
The variational autoencoder is therefore able to learn a ``smooth'' latent space mapping to realistic examples
which provides interesting ways to control the variation of the generation.

\item[\bf Value encoding\index{Value encoding}]
The direct encoding of a numerical value as a scalar.

\item[\bf Vanishing or exploding gradient problem\index{Vanishing gradient problem}\index{Exploding gradient problem}]
A known problem when training a recurrent neural network 
caused by the difficulty of estimating gradients,
because,
in backpropagation through time,
recurrence brings repetitive multiplications
and could thus lead to over amplify or minimize effects (numerical errors).
The long short-term memory (LSTM) architecture solved the problem.


\item[\bf Waveform\index{Waveform}]
The raw representation of a signal as the evolution of its amplitude in time.

\item[\bf Weight\index{Weight}]
A numerical parameter associated to a connexion between a node (neuron or not) and a unit (neuron).
A neuron will compute the weighted sum of the activations of its connexions and then apply its associated activation function.
Weights will be adjusted during the training phase.


\end{description}

\end{document}